\documentclass[preprint,prd,superscriptaddress,showpacs,nofootinbib,floatfix,preprintnumbers]{revtex4-1}
\usepackage{graphicx, bm,color}
\newcommand{\lsim}{\mbox{\raisebox{-.6ex}{~$\stackrel{<}{\sim}$~}}}

\begin{document}
\preprint{UMD-PP-013-011, MAN/HEP/2013/21}
\title{Natural TeV-Scale Left-Right Seesaw for Neutrinos and Experimental Tests}

\author{Chang-Hun Lee} 
\affiliation{Maryland Center for Fundamental Physics and Department of Physics, University of Maryland, College Park, Maryland 20742, USA}

\author{P. S. Bhupal Dev}
\affiliation{Consortium for Fundamental Physics, School of Physics and Astronomy, University of Manchester, Manchester M13 9PL, United Kingdom}

\author{R. N. Mohapatra} 
\affiliation{Maryland Center for Fundamental Physics and Department of Physics, University of Maryland, College Park, Maryland 20742, USA}

\begin{abstract}
We present a TeV-scale left-right ultraviolet completion of type-I seesaw for neutrino masses  based on the $SU(2)_L\times SU(2)_R\times U(1)_{B-L}$ gauge group without parity,  which leads to ``large'' light-heavy neutrino mixing while keeping the neutrino masses small in a natural manner guaranteed by discrete symmetries.  We point out specific observable implications of this class of models if the $SU(2)_R$-breaking scale is of order 5 TeV, in searches for lepton flavor violating processes such as $\mu\to e\gamma$, $\mu\to 3 e$ and $\mu\to e$ conversion in nuclei, and lepton number violating processes such as neutrinoless double beta decay as well as at the LHC.  In particular, if the upper limit on BR$(\mu\to e\gamma)$ improves by one order of magnitude,  a large range of the parameters of the model would be ruled out.
\end{abstract}

\maketitle

\section{Introduction}
The observation of nonzero neutrino masses and mixing has provided the first experimental evidence for physics beyond the Standard Model (SM).  Since the origin of mass for all charged fermions in the SM appears to have been clarified by the discovery of the Higgs boson with mass of 125 GeV at the LHC~\cite{Higgs}, an important question is whether the same Higgs field is also responsible for neutrino masses. If we simply add three right-handed (RH) neutrinos ($\nu_R$)  to the SM, a Yukawa coupling of the form ${\cal L}_{\nu,Y}=h_{\nu,ij} \bar{L}_{i}\Phi \nu_{R,j}+{\rm h.c.}$ can be written,  where $L_i=(\nu_i, \ell_i)_L^{\sf T}$ (with $i=e,\mu,\tau$) is the $SU(2)_L$ lepton doublet, and $\Phi=(\phi^0,\phi^-)^{\sf T}$ is the SM Higgs doublet. After spontaneous symmetry breaking of the electroweak gauge group $SU(2)_L\times U(1)_Y$ to $U(1)_{\rm em}$ via the Higgs vacuum expectation value (vev) $\langle \Phi \rangle = (v,0)^{\sf T}$, this Yukawa term gives masses of the form $h_\nu v$ to the neutrinos. However, to get sub-eV neutrino masses as observed, it requires $h_\nu \lsim 10^{-12}$ which is an unnaturally small number. This provides sufficient reason to believe that there is some other new physics behind neutrino masses, beyond adding just three RH neutrinos to the SM, thereby providing the first clue to the nature of physics beyond the SM. 

A simple paradigm for understanding the small neutrino masses is the type-I seesaw mechanism~\cite{type1} where the RH neutrinos alluded to above have a Majorana mass of the form $M_N\nu_R^{\sf T}\nu_R$,  in addition to having  Dirac masses like all charged fermions in the SM. Neutrinos being electrically neutral allow for this possibility, distinguishing them from the charged fermions, and this feature might be at the heart of such diverse mass and mixing patterns for leptons in contrast with the quark sector.  The seesaw mechanism leads to the following generic $6\times 6$ neutrino mass matrix in the $\{\nu_L^C, \nu_R\}$ flavor basis ($C$ being the charge conjugation):
\begin{eqnarray}
{\cal M}_\nu = \left(\begin{array}{cc}0 & M_D\\ M^{\sf T}_D & M_N\end{array}\right) 
\label{seesaw}
\end{eqnarray}
where the $3\times 3$ Dirac mass matrix $M_D$ mixes the $\nu_L$ and $\nu_R$ states (the later also generically denoted by $N$) and is generated by the SM Higgs field, while $M_N$ is the Majorana mass for $\nu_R$ which embodies the new neutrino mass physics. In the usual seesaw approximation $\|\xi\| \ll 1$, where the heavy-light neutrino mixing matrix $\xi\sim M_DM_N^{-1}$ and $\|\xi\|\equiv  \sqrt{{\rm Tr} (\xi^\dagger \xi)}$, the light neutrino mass matrix is given by the seesaw formula~\cite{type1}
\begin{eqnarray}
M_\nu \simeq -M_DM_N^{-1}M_D^{\sf T}.
\label{seesaw2}
\end{eqnarray}
Our goal in this paper  is to explore whether the two key aspects of this new seesaw physics, i.e., (i) the Majorana character of heavy and light neutrino masses, and (ii) the heavy-light neutrino mixing, can be tested at the LHC  as well as in complementary experiments at low energies, e.g.,~in planned high sensitivity searches for charged lepton flavor violation (LFV), non-unitarity of the leptonic mixing matrix, neutrinoless double beta decay ($0\nu\beta\beta$), etc.
 A necessary requirement for this synergic exploration to have any chance of success is that  the seesaw scale be in the TeV range as well as the heavy-light mixing being relatively large. With this in mind, we discuss in this article a class of models where both the above ingredients of 
type-I seesaw, i.e., TeV seesaw scale and observable heavy-light neutrino mixing emerge in a natural manner. 

A simple candidate for an ultraviolet (UV)-complete seesaw model is based on the left-right (L-R) symmetric   theory of weak interactions, based on the gauge group $SU(2)_L\times SU(2)_R\times U(1)_{B-L}$~\cite{LR}, where the key ingredients of seesaw, i.e., the RH  neutrino and its Majorana mass, appear naturally.  The RH neutrino field $\nu_R$ arises as the necessary parity gauge partner of the left-handed (LH) neutrino field $\nu_L$ and is 
also required by anomaly cancellation, whereas the seesaw scale is identified as the one at which the  RH counterpart of the SM $SU(2)_L$ gauge symmetry, namely the $SU(2)_R$ symmetry, is broken. 
The RH neutrinos are therefore a necessary part of the model and do not have to be added just to implement the seesaw mechanism. An important point is that the RH neutrinos acquire a Majorana mass as soon as the $SU(2)_R$ symmetry is broken at a scale $v_R$. This is quite analogous to the way the charged fermions get mass as soon as the SM gauge symmetry $SU(2)_L$ is broken at the electroweak scale $v$. The Higgs field that gives mass to the RH neutrinos becomes the analog of the 125 GeV Higgs boson discovered at the LHC. Clearly, the seesaw scale is not added in an adhoc manner but rather becomes intimately connected to the $SU(2)_R\times U(1)_{B-L}$ symmetry breaking scale. 
 
 Since L-R symmetric theories lead to new effects or add new contributions to already known low energy weak processes, it is necessary to know whether TeV scale for $SU(2)_R$-breaking is compatible with observations of low energy processes. It turns out that hadronic flavor changing neutral current (FCNC) effects such as $\epsilon_{K}$ and $K_L-K_S,~B_S-\overline{B}_S$ mixing receive significant contributions from RH charged current effects, and therefore, provide stringent constraints on the $SU(2)_R$-breaking scale $v_R$ by restricting the mass of the RH charged $W_R$ boson  to be $M_{W_R} \geq 2.5$ TeV~\cite{KL-KS}. The direct search 
 limits on $M_{W_R}$ from the $\sqrt s=7$ and 8 TeV LHC data are currently between 2.5 - 2.9 TeV~\cite{ATLAS-RR}. At $\sqrt s=14$ TeV LHC, the L-R  model as a theory of neutrino mass can be probed as long as $M_{W_R}$ is below 6 TeV~\cite{ferrari}. There are also low energy complementary tests of the L-R model in the domain of leptonic physics (for a recent review, see Ref.~\cite{deGouvea:2013zba}), e.g., lepton flavor violating processes such as $\mu\to e\gamma$, $\mu\to 3e$, $\mu\to e$ conversion in nuclei, and lepton number violating  processes such as $0\nu\beta\beta$. 

In generic TeV-scale seesaw models without any special structures for $M_D$ and $M_N$, in order to get small neutrino masses, we must fine-tune the magnitude of the elements of $M_D$ to be very small (of order MeV for $M_N\sim$ TeV), as is evident from the seesaw formula in  Eq.~(\ref{seesaw2}). As a result, the heavy-light neutrino mixing $\xi\sim M_DM_N^{-1}\simeq (M_\nu M_N^{-1})^{1/2}\lsim 10^{-6}$.~\footnote{Note that this is true for a {\it generic} type-I seesaw model regardless of whether the seesaw scale is in the TeV range or higher.} This suppresses all heavy-light mixing effects to an unobservable level which keeps this key aspect of seesaw shielded from being tested experimentally. To overcome this shortcoming, some special textures for $M_D$ and $M_N$ have been studied in the literature~\cite{cancel, cancel1} for which even with TeV-scale seesaw, the mixing parameter $\xi$ can be significantly enhanced whereas the neutrino masses still remain small, thereby enriching the seesaw phenomenology.  In this paper, we present an L-R model embedding of one such special texture using an appropriate family symmetry. This is a highly non-trivial result since in L-R models the charged lepton mass matrix and the Dirac neutrino mass matrix are related, especially when there are additional discrete symmetries to guarantee a specific form of the Dirac mass matrix $M_D$, as discussed below.  After obtaining realistic fits to the observed charged lepton masses, and light neutrino masses and mixing, while simultaneously satisfying all the other existing experimental constraints in the leptonic sector, we study the phenomenological implications of our solutions for collider signals, non-unitarity of the leptonic mixing matrix, LFV effects and $0\nu\beta\beta$ processes which could be used to test this model. We present results for the specific case of $v_R= 5$ TeV but our results can easily be scaled accordingly for different $v_R$ values. Any evidence for the considerations of this paper can reveal underlying symmetries of the lepton sector, which will be an important step towards a full understanding of the seesaw mechanism.
 
 The rest of the article is organized as follows: in Section~\ref{sec2}, we review the basic features of the generic L-R seesaw models. In Section~\ref{sec3}, we present an L-R seesaw model where a discrete leptonic family symmetry leads to  special Dirac and Majorana textures resulting in light neutrinos and `large'  heavy-light neutrino mixing via type-I seesaw in a natural manner.  In Section~\ref{sec4}, we present our numerical fit results for the model parameters. In Section~\ref{sec5}, we discuss the implications of our results for lepton flavor violation. In Section~\ref{sec6}, we give the model predictions for large heavy-light mixing effects in the light neutrino mixing matrix. In Section~\ref{sec7}, we discuss the model implications for neutrinoless double beta decay. In Section~\ref{sec8}, we discuss the collider signatures of this model.  Our conclusions are given in Section~\ref{sec9}.
\section{ Review of the generic left-right seesaw model}\label{sec2}
Before discussing the details of our model, we review, for completeness, the basic features of generic L-R seesaw models. In the minimal L-R model, the fermions are assigned to the gauge group $SU(2)_L\times SU(2)_R\times U(1)_{B-L}$ as follows: denoting 
$Q\equiv (u,d)^{\sf T}$ and $\psi\equiv (\nu_\ell, \ell)^{\sf T}$ as the quark and lepton doublets respectively, $Q_L$ and $\psi_L$ (also denoted simply by $L$) are assigned to doublets under the $SU(2)_L$ group, while $Q_R$ and $\psi_R$ (also denoted by $R$) doublets under the $SU(2)_R$ group. Their $B-L$ quantum numbers can easily be worked out from the definition of the electric charge: $Q=I_{3L}+I_{3R}+(B-L)/2$, where $I_{3L}$ and $I_{3R}$ are the third components of isospin under $SU(2)_L$ and $SU(2)_R$ respectively.  
The Higgs sector of the model consists of one or several of the following multiplets:
\begin{eqnarray}
\Delta_R\equiv\left(\begin{array}{cc}\Delta^+_R/\sqrt{2} & \Delta^{++}_R\\\Delta^0_R & -\Delta^+_R/\sqrt{2}\end{array}\right),
\quad \phi\equiv\left(\begin{array}{cc}\phi^0_1 & \phi^+_2\\\phi^-_1 & \phi^0_2\end{array}\right).
\end{eqnarray}
The gauge symmetry $SU(2)_R\times U(1)_{B-L}$ is broken by the vev $\langle \Delta^0_R\rangle = v_R$ to the group $U(1)_Y$ of the SM. 
There is also an LH counterpart $\Delta_L$, which we do not consider here for the following reasons.  There are versions of the model where parity and $SU(2)_R$ gauge symmetry scales are decoupled so that the $\Delta_L$ fields become heavy when the discrete parity symmetry is broken, and disappear from the low energy theory~\cite{CMP}. This version (i.e., L-R model without parity) is also necessary to suppress the type-II contribution to neutrino mass in low scale L-R models~\cite{chang}, which otherwise could be unacceptably large. The low-energy Lagrangian in this case therefore has invariance under the L-R gauge group but not parity. We will focus on this class of models in this paper, which seem to be necessary in implementing our strategy. 

In generic models, the vev of the $\phi$ field given by $\langle\phi\rangle={\rm diag}(\kappa, \kappa')$ breaks the SM gauge group to $U(1)_{\rm em}$.
To see how the fermions pick up mass and how the seesaw mechanism arises in the L-R model, we write down the Yukawa Lagrangian of the model:
\begin{eqnarray}
{\cal L}_Y&=&h^{q,a}_{ij}\bar{Q}_{L,i}\phi_aQ_{R,j}+\tilde{h}^{q,a}_{ij}\bar{Q}_{L,i}\tilde{\phi}_aQ_{R,j}+
h^{\ell,a}_{ij}\bar{L}_i\phi_aR_j \nonumber \\
&&\qquad
+~\tilde{h}^{\ell,a}_{ij}\bar{L}_i\tilde{\phi}_aR_j
+ f_{ij} (R_iR_j\Delta_R + L_iL_j\Delta_L)+{\rm h.c.},
\label{eq:yuk}
\end{eqnarray}
where $i,j$ stand for generations and $a$ for labeling the Higgs bi-doublets, and $\tilde{\phi}=\tau_2\phi^*\tau_2$ ($\tau_2$ being the second Pauli matrix). After symmetry breaking, the Dirac fermion masses are given by the generic formula $M_f = h^f\kappa + \tilde{h}^f\kappa'$ for up-type fermions, and for down-type quarks and charged leptons, it is the same formula with $\kappa$ and $\kappa'$ interchanged.  The Yukawa Lagrangian in Eq.~(\ref{eq:yuk}) leads to the Dirac mass matrix  for neutrinos $M_D = h^{\ell}\kappa + \tilde{h}^{\ell}\kappa'$ and the Majorana mass matrix for the heavy RH  neutrinos $M_N = fv_R$ which go into Eq.~(\ref{seesaw}) for calculating the neutrino masses and the heavy-light neutrino mixing.

As discussed earlier, for generic forms of $M_D$ and $M_N$, the heavy-light mixing parameter $\xi \simeq [M_\nu(fv_R)^{-1}]^{1/2}$ is a tiny number regardless of whether the seesaw scale is in the TeV range or higher. This keeps all heavy-light mixing effects at an unobservable level. In the following section, we present a model with special textures for $M_D$ and $M_N$ guaranteed by an appropriate discrete symmetry which leads to large heavy-light neutrino mixing with TeV-scale type-I seesaw in a natural way, thereby enhancing the phenomenological richness of the model.
\section{Enhanced Heavy-light neutrino mixing with special Dirac and Majorana textures}\label{sec3}
\subsection{Basic strategy}
The basic strategy for implementing our program is to have the appropriate textures for $M_D$ and $M_N$ which via type-I seesaw lead to `large' heavy-light mixing and observed light neutrino masses. There are several examples of this type 
discussed in the literature~\cite{cancel, cancel1}. In this section, we discuss the embedding of one of them (from Ref.~\cite{cancel}) in the L-R model using an appropriate family symmetry. The symmetry must not only guarantee the special leptonic textures but also must be free of light scalar bosons which can result if the effect of the discrete symmetry is to automatically lead to a $U(1)$ symmetry of the full Lagrangian. Moreover in L-R symmetric models, the charged lepton mass matrix can be related to $M_D$ which puts additional constraints on the phenomenological viability of the model. We therefore find it remarkable that the model we present below remains a viable TeV-scale L-R type-I seesaw model for neutrinos, and as a result, has interesting phenomenological implications. 

The Dirac and Majorana mass matrices $M_D$ and $M_N$ considered here have the following form:
\begin{eqnarray}
M_D=\left(\begin{array}{ccc} m_1 & \delta_1 & \epsilon_1\\ m_2 & \delta_2 & \epsilon_2\\ m_3 &\delta_3 & \epsilon_3\end{array}\right), \quad
M_N=\left(\begin{array}{ccc} 0 & M_1 & 0\\M_1 & \delta M & 0 \\ 0&0&M_2\end{array}\right) \label{eq:texture}
\end{eqnarray}
with $\epsilon_i, \delta_i \ll m_i $ and $\delta M \ll M_i$. In the limit of $\epsilon_i, \delta_i, \delta M\to 0$, the neutrino masses vanish, although the heavy-light mixing parameters given by $\xi_{ij}\sim m_i/M_j$ (roughly speaking) can be quite large. The neutrino masses given by the seesaw formula are dependent upon the small parameters $\epsilon_i$ and $\delta_i$. If by some symmetry one can guarantee the smallness of $\delta_i$ and $\epsilon_i$, then we have a TeV-scale seesaw model with enhanced heavy-light mixing. As we show below, these mass textures can be embedded into L-R models and can reproduce the observed neutrino masses and mixing without fine-tuning of parameters. Since in L-R models the charged lepton mass matrix and the Dirac neutrino mass matrix are related, fitting neutrino oscillation data in a manner that also reproduces charged lepton masses is specially nontrivial, in addition to the fact that the seesaw scale is in TeV range.  
\subsection{Naturalness of the special Dirac and Majorana textures} 
In order to obtain the special Dirac and Majorana textures given in Eq.~(\ref{eq:texture}), we use only the L-R gauge group $SU(2)_L\times SU(2)_R\times U(1)_{B-L}$ without the parity symmetry and  supplement it with a global discrete symmetry $D\equiv Z_4\times Z_4\times Z_4$. For the Higgs sector, we choose three bi-doublets ($\phi_{1,2,3}$) with $B-L=0$ and two RH triplets ($\Delta_{R1, R2}$) with $B-L=2$. The fermion and Higgs multiplets are assigned the $D$ quantum numbers as shown in Table~\ref{tab1}. 
\begin{table}[t]
\begin{center}
\begin{tabular}{c|c}\hline\hline
Field & $Z_4\times Z_4\times Z_4$ Transformation\\ \hline
$L_\alpha$ & $(1,\ 1,\ 1)$\\
$R_1$ & $(-{\rm i},\ 1,\ 1)$ \\
$R_2$ & $(1,\ -{\rm i},\ 1)$ \\
$R_3$ & $(1,\ 1,\ -{\rm i})$ \\
$\phi_1$ & $(-{\rm i},\ 1,\ 1)$\\
$\phi_2$ & $(1,\ {\rm i},\ 1)$\\
$\phi_3$ & $(1,\ 1,\ {\rm i})$\\
$\Delta_{R,1}$ & $({\rm i}, \ {\rm i},\ 1)$\\
$\Delta_{R,2}$ & $(1, \ 1,\ -1)$\\
\hline\hline
\end{tabular}
\end{center}
\caption{The discrete symmetry assignments for the fermion and Higgs fields in our L-R model that lead naturally to the special Dirac and Majorana textures given in Eq.~(\ref{eq:texture}).  Here $\alpha=1,2,3$ for different generations, and i$\equiv \sqrt{-1}$.}\label{tab1}
\end{table}

The leptonic Yukawa Lagrangian invariant under this symmetry $D$ is given by
\begin{eqnarray}
{\cal L}_{\ell,Y}=h_{\alpha 1}\bar{L}_\alpha \tilde{\phi}_1R_1+h_{\alpha 2} \bar{L}_\alpha\phi_2R_2+h_{\alpha 3}\bar{L}_\alpha\phi_3R_3
+f_{12}R_1R_2\Delta_{R,1}+f_{33}R_3R_3\Delta_{R,2}+{\rm h.c.}
\label{eq:L_l}
\end{eqnarray}
An important point to emphasize is that  in the discrete symmetry limit, the vevs of $\phi_{1,2,3}$ will have the following form:
\begin{eqnarray}
\langle \phi_{1,2,3}\rangle = \left(\begin{array}{cc} 0 & 0 \\0 & \kappa'_{1,2,3}\end{array}\right).
\end{eqnarray}
Note that there is a range of parameters where this minimum is stable. An easy way to see this is to note that  terms of the form Tr$ (\tilde{\phi}_{a}\phi^\dagger_{b})$ which would change the $\phi$ vev to the form diag$ (\kappa, \kappa')$ are forbidden from appearing in the scalar potential due to the discrete symmetry.
So we see that in the symmetry limit, the Dirac mass matrix $M_D$ has large and nonzero first column entries, and the charged lepton mass matrix has one eigenvalue zero which we can identify as the electron flavor. 
To make the model realistic, we add very tiny soft symmetry-breaking terms to the scalar potential as follows:
\begin{eqnarray}
\delta V(\phi)= \sum_{a,b=1}^3\mu^2_{ab}{\rm Tr}(\tilde{\phi}_{a}\phi^\dagger_b) + {\rm h.c.}
\end{eqnarray}
Note that $\mu^2_{ab}$ are only multiplicatively renormalized. As a result, we can choose their magnitudes to be very small without worrying about radiative destabilization. This will then cause the $\phi_{a}$ vevs to take the following form:
\begin{eqnarray}
\langle \phi_{1,2,3}\rangle &=& \left(\begin{array}{cc} \delta \kappa_{1,2,3} & 0 \\0 & \kappa'_{1,2,3}\end{array}\right),\\
{\rm where}~~ \delta \kappa_a &\propto& \frac{\displaystyle \sum_{b=1}^3 \mu^2_{ab}\kappa'_b}{\displaystyle \sum_{a,b=1}^3\lambda'_{ab}\kappa^\prime_a\kappa^\prime_b + \displaystyle \sum_{a=1}^2\lambda_a v_{R,a}^2},\label{eq:delka}
\end{eqnarray}
and $\lambda'$, $\lambda$ are respectively the generic scalar self couplings of the bi-doublet and triplet fields in the scalar potential. 
Choosing $\mu^2_{ab}$ to be appropriately small, we can get very small $\delta \kappa_{a}$ that we use in the mass fits below. It is possible to generate these small numbers naturally through loop effects involving the $W_L-W_R$ mixing or from higher dimensional terms in the potential using gauge singlet but $Z_4\times Z_4 \times Z_4$ non-singlet scalar fields which acquire vev at a scale higher than $v_R$.

An important point to notice is that the $\delta \kappa_a$'s defined in Eq.~(\ref{eq:delka}) are responsible for the electron mass as well as neutrino masses via type-I seesaw. Thus getting a fit to the observed neutrino masses and mixing while at the same time keeping electron mass at its desired value is a nontrivial task since in the lepton sector, the model has only 12 free parameters, \footnote{Even though a naive counting of parameters in Eq.~(\ref{eq:L_l}) suggests the number of parameters to be 15, one can rotate the LH lepton doublets before symmetry breaking, and reduce the number of Yukawa parameters to six. The final fit remains unchanged due to this unitary rotation. After submitting this paper for publication, we have found solutions where $\delta M=0$ in Eq.~(\ref{eq:texture}) in which case the total number of free parameters actually reduces to 11.} and out of this, we must not only get fits for the three charged lepton masses, the two neutrino mass-squared differences and three mixing angles (total of 8 outputs), but we must also satisfy the unitarity constraints on the new light neutrino mixing matrix as well as constraints from rare lepton decays which involve only the RH mixing matrix for charged leptons.
There are therefore a lot more constraints than inputs and
below we show that our model does indeed provide a fit to all observables while satisfying all the necessary constraints.

The following points about this model are worth emphasizing:

\begin{itemize}

\item  There is another bi-doublet field $\phi_0$ in the model  which is a singlet under the discrete group $D$ and does not couple to leptons but rather is used to give mass to the quarks (which are also assumed to be singlets under $D$).  Its complex Yukawa couplings can then generate the CKM mixing angles as well as the phase for $C\!P$ violation in the quark sector. It is also responsible for mixing between the $W_L$ and $W_R$ gauge fields which has implications for $0\nu\beta\beta$ and the electric dipole moment of the neutron~\cite{Nemevsek}. An alternative possibility is that quark sector has no $\phi_0$ field but rather a quark seesaw so that quarks get their mass from heavy, iso-singlet  vector-like quarks. In this case, the small $\delta \kappa$ values needed to give neutrino masses can actually be predicted. A model of this type is now under investigation. 

\item It appears that the choice of the product of $Z_4$ groups in Table~\ref{tab1} reduces possible multiple $U(1)$ symmetries of the model associated with different bi-doublets since it allows terms like Tr$[(\phi^\dagger\tilde{\phi})]^2$ in the Higgs potential. One may also add soft $D$-breaking terms of the form
${\rm Tr}(\phi^\dagger_a{\phi}_b)$ ($a,b=1,2,3)$ in the potential which do not destabilize the minimum we are interested in. The detailed analysis of the Higgs potential and its phenomenology will be the subject of a future publication. 

\item  A consequence of a separate leptonic family symmetry $D$ is that our model predicts sub-TeV SM-like heavy Higgs bosons which are leptophilic. They arise from the three bi-doublets $\phi_{1,2,3}$ which couple only to leptons. Their masses are in the sub-TeV range due to the fact that they do not couple to the RH triplets $\Delta_{1,2}$ in the Higgs potential and participate in the SM symmetry breaking. The detailed phenomenology of these Higgs bosons will be discussed elsewhere, but the general point worth making is that if the LHC limits on SM-like but leptophilic Higgs bosons exceed $\sim$ TeV, our model will be under serious tension.

\item In our discussion, we have ignored the $C\!P$ phases in the  lepton Yukawa couplings. There are three independent $C\!P$ phases, after redefinition of the lepton fields. We are currently investigating whether there are solutions with nontrivial $C\!P$ phases. This issue is clearly very important~\cite{lepCP} in view of the general belief that it may hold the clue to understanding the origin of matter in the universe as well as that one of the major goals of the next generation of neutrino experiments is to search for any possible leptonic $C\!P$ violation.
\end{itemize}
\section{Neutrino mass fits and range of parameters in the model}\label{sec4}
In this section, we present the best-fit as well as the ranges of parameters that produce the measured values of the charged lepton masses, the neutrino mass-squared differences, and the lepton mixing angles. After all the neutral scalar fields acquire their vevs, the charged lepton mass matrix, and the Dirac and Majorana neutrino mass matrices are respectively given by
\begin{eqnarray}
	M_\ell &=&
	\left(\begin{array}{ccc}
		h_{11} \delta \kappa_1 & h_{12} \kappa'_2 & h_{13} \kappa'_3 \\
		h_{21} \delta \kappa_1 & h_{22} \kappa'_2 & h_{23} \kappa'_3 \\
		h_{31} \delta \kappa_1 & h_{32} \kappa'_2 & h_{33} \kappa'_3
	\end{array} \right), \label{eq:Mll} \\
	M_D &=&
	\left(\begin{array}{ccc}
		h_{11} \kappa_1' & h_{12} \delta \kappa_2 & h_{13} \delta \kappa_3 \\
		h_{21} \kappa_1' & h_{22} \delta \kappa_2 & h_{23} \delta \kappa_3 \\
		h_{31} \kappa_1' & h_{32} \delta \kappa_2 & h_{33} \delta \kappa_3
	\end{array} \right), \label{eq:MD} \\
	M_N &=&
	\left(\begin{array}{ccc}
		0 & f_{12} v_{R,1} & 0 \\
		f_{12} v_{R,1} & \delta M & 0 \\
		0 & 0 & 2 f_{33} v_{R,2}
	\end{array} \right). \label{eq:MN}
\end{eqnarray}
For numerical purposes, we choose a specific value of the $SU(2)_R$-breaking scale $v_R\sim 5$ TeV (compatible with the current limits on $M_{W_R}$) but our results can easily be scaled accordingly for higher values of $v_R$. We further assume the following: (i) each column of the charged lepton mass matrix is of the order of the corresponding lepton mass, i.e.~$h_{\alpha 1} \delta \kappa_1 \sim 0.5 ~ \text{MeV}, \ h_{\alpha 2} \kappa_2' \sim 100 ~ \text{MeV}, \ h_{\alpha 3} \kappa_3' \sim 1 ~ \text{GeV}$, (ii)  the heavy neutrino masses $M_{1,2}$ are in the TeV range, where $M_1\equiv f_{12}v_{R,1}$ and $M_2\equiv 2f_{33}v_{R,2}$, and (iii) $m_i$ are in the GeV range, which is a necessary condition to have large  enough mixing $\xi \sim M_D M_N^{-1}$. Applying the seesaw formula in Eq.~(\ref{seesaw2}) to the mass textures in Eq.~(\ref{eq:texture}), we can write the (3, 3) component of the light neutrino mass matrix as
\begin{eqnarray}
	M_{\nu, 33} \approx -\frac{\epsilon_3^2}{M_2} - \frac{2 m_3 \delta_3}{M_1} + \frac{m_3^2 \delta M}{M_1^2},
	\label{eq:nu33}
\end{eqnarray}
where $\epsilon_3 = h_{33} \delta \kappa_3$, $m_3 = h_{31} \kappa_3'$, and $\delta_3 = h_{32} \delta \kappa_2$ according to Eq.~(\ref{eq:MD}). Since the largest neutrino mass is around $\sqrt{\Delta m^2_{\rm atm}}\sim 0.05$ eV for a normal hierarchy, it is natural to assume that each term in Eq.~(\ref{eq:nu33}) is of the order of $10^{-2}$ eV. For $M_2 \sim 1$ TeV, the condition $\epsilon_3^2 / M_2 \sim 10^{-2}$ eV implies $\epsilon_3 = h_{33} \delta \kappa_3 \sim 100$ eV, which in turn means $\kappa_3' / \delta \kappa_3 \sim 10^4$ for $h_{33} \kappa_3' \sim 1$ GeV. Similarly, for $m_3 \sim$ 1 - 10 GeV and $M_1 \sim$ 0.1 - 1 TeV, 
Eq.~(\ref{eq:nu33}) implies that $\kappa_2' / \delta \kappa_2 \sim 10^7$ - $10^9$ and $\delta M \sim 10^{-9}$ - $10^{-5}$ GeV. Furthermore, $m_3 = h_{31} \kappa_1 \sim$ 1 - 10 GeV and $h_{31} \delta \kappa_1 \sim$ 0.1 - 1 MeV imply $\kappa_1'/ \delta \kappa_1 \sim 10^3$ - $10^5$. By randomly generating parameters around these expected ranges and diagonalizing the charged lepton mass matrix given by Eq.~(\ref{eq:Mll}) and the $6 \times 6$ neutrino mass matrix ${\cal M}_{\nu}$ given by Eq.~(\ref{seesaw}), 
we can obtain the mass eigenvalues for the charged leptons as well as light neutrinos, and the mixing angles in the lepton sector. The best-fit parameter values are given in Table~\ref{tab2}, and  the corresponding mass and mixing parameters are given in Table~\ref{tab3}.
The mixing matrices for the LH charged lepton and the light neutrinos corresponding to these best-fit model parameters are respectively given by
\begin{eqnarray}
	V_\ell^L &=&
	\left(\begin{array}{ccc}
		0.780 & -0.285 & 0.557 \\
		-0.341 & -0.940 & -0.00318 \\
		0.525 & -0.188 & -0.830
	\end{array} \right), \label{eq:VeL} \\
	V_\nu^L &=&
	\left(\begin{array}{ccc}
		0.991 & -0.0197 & 0.131 \\
		0.0544 & -0.842 & -0.536 \\
		0.121 & 0.539 & -0.834
	\end{array} \right),
\label{eq:VnuL}
\end{eqnarray}
from which we obtain
\begin{eqnarray}
	V_{\text{PMNS}} &\equiv (V_\ell^L)^\dagger V_\nu^L 
	= \left(\begin{array}{ccc}
		0.818 & 0.555 & -0.152 \\
		-0.356 & 0.696 & 0.623 \\
		0.452 & -0.455 & 0.767
	\end{array} \right).
\label{eq:pmns}
\end{eqnarray}
Here, we have taken the Dirac $C\!P$ phase in the neutrino sector to be $\pi$, and the Majorana phases to be zero for illustration. Comparing the output parameters in Table III with the current experimental values~\cite{PDG}, we find that the charged lepton masses are off from the measured values by at most 1\% which can be easily accounted for due to electroweak radiative corrections. The solar and atmospheric neutrino mass-squared differences and the three mixing angles are within their $1\sigma$ global fit values~\cite{gonzalez}.   
\begin{table}[t]
\begin{center}
\begin{tabular}{c|c} \hline\hline
Input Parameter & Best-Fit Value \\ \hline
$h_{11} \delta \kappa_1$ & $2.63 \times 10^{-4}$ GeV \\
$h_{21} \delta \kappa_1$ & $-1.99 \times 10^{-4}$ GeV \\
$h_{31} \delta \kappa_1$ & $-1.49 \times 10^{-3}$ GeV \\
$h_{12} \kappa_2'$ & $0.0151$ GeV \\
$h_{22} \kappa_2'$ & $0.0992$ GeV \\
$h_{32} \kappa_2'$ & $0.0421$ GeV \\
$h_{13} \kappa_3'$ & $-0.989$ GeV \\
$h_{23} \kappa_3'$ & $0.0413$ GeV \\
$h_{33} \kappa_3'$ & $1.47$ GeV \\
$\kappa_1'/\delta \kappa_1$ & $9.14 \times 10^3$ \\
$\kappa_2' / \delta \kappa_2$ & $6.76 \times 10^7$\\
$\kappa_3' / \delta \kappa_3$ & $9.11 \times 10^3$\\
$f_{12} v_{R, 1}$ & $813$ GeV \\
$f_{33} v_{R, 2}$ & $-2490$ GeV \\
$\delta M$ & $1.30 \times 10^{-9}$ GeV \\
\hline\hline
\end{tabular}
\end{center}
\caption{The best-fit values of the parameters in our L-R model.}
\label{tab2}
\end{table}
\begin{table}[t]
\begin{center}
\begin{tabular}{c|c} \hline \hline
Output Parameter & Value \\ \hline
$m_e$ & $0.511$ MeV  \\
$m_\mu$ & $105.52$ MeV \\
$m_\tau$ & $1.77$ GeV \\ \hline
$m_{\nu_1}$ & $2.91 \times 10^{-3}$ eV \\
$m_{\nu_2}$ & $-9.18 \times 10^{-3}$ eV \\
$m_{\nu_3}$ & $4.96 \times 10^{-2}$ eV \\
$m_{N_1}$ & $-813.37$ GeV \\
$m_{N_2}$ & $813.37$ GeV \\
$m_{N_3}$ & $-2488.96$ GeV \\ \hline
$\theta_{12}$ & $34.15^\circ$ \\
$\theta_{23}$ & $39.08^\circ$ \\
$\theta_{13}$ & $8.74^\circ$ \\ \hline
\hline
\end{tabular}
\end{center}
\caption{The best-fit values of the lepton masses and mixing corresponding to the model parameter values given in Table~\ref{tab2}.}
\label{tab3}
\end{table}

Since it is convenient to evaluate the low-energy observables in the leptonic sector (discussed in the following sections) in a basis where the charged lepton mass matrix is diagonal, we perform the following transformations to change to this basis from the one defined by Eqs.~(\ref{eq:Mll}) - (\ref{eq:MN}): 
\begin{eqnarray}
	M_D&\to& M_D' = (V_\ell^L)^\dagger M_D V_\ell^R,\nonumber\\
	M_N&\to& M_N' = (V_\ell^R)^{\sf T} M_N V_\ell^R, 
\end{eqnarray}
where $V_\ell^{L,R}$ are the mixing matrices for the LH and RH charged leptons respectively. With this redefinition, we diagonalize the full neutrino mass matrix ${\cal M}_\nu$ given by Eq.~(\ref{seesaw}) by a $6\times 6$ unitary matrix
 \begin{eqnarray}
	{\cal V}_\nu &\equiv & \left(\begin{array}{cc}
	U_L & \xi \\
	\xi' & V_R
	\end{array}\right),
\label{eq:big}
\end{eqnarray}
 whose upper $3\times 3$ block $U_L$ will represent the new PMNS mixing matrix diagonalizing the light neutrinos, while the off-diagonal $3\times 3$ blocks represent the heavy-light neutrino mixing matrices $\xi$ and $\xi'$ which are crucial ingredients for our model phenomenology.  In our model, the RH charged lepton mixing matrix $V_{\ell}^R$ is close to being the identity matrix, with the off-diagonal (1,2) element of the order of $10^{-3}$ which is small enough to suppress the $\mu\to 3e$ LFV process due to contributions from the $\Delta_R$ fields (see Section~\ref{sec4}). The best-fit values for $V_\ell^L$ and $U_L$ are already given in Eqs.~(\ref{eq:VeL}) and (\ref{eq:pmns}) respectively. The corresponding best-fit values for the remaining mixing matrices are given below: 
\begin{eqnarray}
V_\ell^R &=& \left(\begin{array}{ccc}
1 & -0.0037 & -0.0008 \\
0.0037 & 0.9999 & 0.0151\\
0.0007 & -0.0151 & 0.9999
\end{array}\right),\\
V_R &=& \left(\begin{array}{ccc}
	-0.7045 & 0.7097 & -0.0007 \\
	0.7095 & 0.7043 & 0.0151\\
	0.0112 & 0.0101 & -0.9999
	 \end{array}\right),\label{eq:VnR} \\
	\xi &=& \left(\begin{array}{ccc}
	 0.0041 & 0.0041 & -8.2\times 10^{-13}\\
	 -0.0031 & -0.0031 & -7.0\times 10^{-11}\\
	 -0.0110 & -0.0110 & 7.8\times 10^{-8}
	 \end{array}\right),\label{eq:bfxi}\\
		 \xi' &=& \left(\begin{array}{ccc}
	 1.8\times 10^{-6} & 2.6\times 10^{-5} & 5.8\times 10^{-5} \\
	4.8\times 10^{-4} & 0.0071 & 0.0157\\
	7.3\times 10^{-6} & 0.0001 & 0.0002
	  \end{array}\right).
\end{eqnarray}
 
In addition to the best-fit values given in Table~\ref{tab2}, we provide in Table~\ref{tab4} a sample set of parameter ranges which can produce the charged lepton masses within 1\% of measured values and the neutrino masses as well as mixing within $3\sigma$ of their global fit values. Note that these parameter ranges are not the broadest ones, and it might be possible to have a different set of parameter ranges still satisfying all the existing data. We also note that by an appropriate scaling of the parameters $\delta \kappa_a/\kappa_a'$ and $v_R$, our  fits for the neutrino oscillation parameters and the charged lepton masses remain unchanged. For definiteness, we focus our attention on parameters for which the lightest RH neutrino mass is less than a TeV so that it has sizable effects at the LHC and other low-energy experiments.
\begin{table}[t]
\begin{center}
\begin{tabular}{c|c} \hline \hline
Model Parameter & Range  \\ \hline
$h_{11} \delta \kappa_3$ & (2 - 4) $\times \ 10^{-4}$ GeV \\
$h_{21} \delta \kappa_3$ & $-$(3 - 1) $\times \ 10^{-4}$ GeV \\
$h_{31} \delta \kappa_3$ & $-$(1.7 - 1.3) $\times \ 10^{-3}$ GeV \\
$h_{12} \kappa_2'$ & (0 - 0.02)  GeV \\
$h_{22} \kappa_2'$ & (0.095 - 0.105) GeV \\
$h_{32} \kappa_2'$ & (0.03 - 0.06) GeV \\
$h_{13} \kappa_3'$ & $-$(1.2 - 0.8) GeV \\
$h_{23} \kappa_3'$ & $-$(0.01 - 0.01) GeV \\
$h_{33} \kappa_3'$ & (1.3 - 1.6) GeV \\
$\kappa_1' / \delta \kappa_1$ & (6 - 10) $\times \ 10^3$ \\
$\kappa_2' / \delta \kappa_2$ & (5 - 7) $\times \ 10^7$  \\
$\kappa_3' / \delta \kappa_3$ & (6 - 10) $\times \ 10^3$ \\
$f_{12} v_{R, 1}$ & (600 - 1000) GeV \\
$f_{33} v_{R, 2}$ & $-$(5000 - 1500) GeV \\
$\delta M$ & (1 - 5) $\times \ 10^{-9}$ GeV \\
\hline\hline
\end{tabular}
\end{center}
\caption{A sample set of model parameter ranges satisfying all experimental constraints.}\label{tab4}
\end{table}
\section{Lepton Flavor Violation consequences}\label{sec5}
In this section, we discuss the implications of our model for LFV.
In the ``vanilla'' (i.e., without any special textures for $M_D$ and $M_N$) as well as the special class of TeV-scale L-R seesaw models being discussed here, the LFV  processes receive new contributions from the RH currents not present in the SM-seesaw (without L-R symmetry), as discussed below: 
\subsection{$\mu\to e\gamma$}
\begin{figure}[t]
\begin{center}
\begin{tabular}{ccc}
\includegraphics[width=5cm]{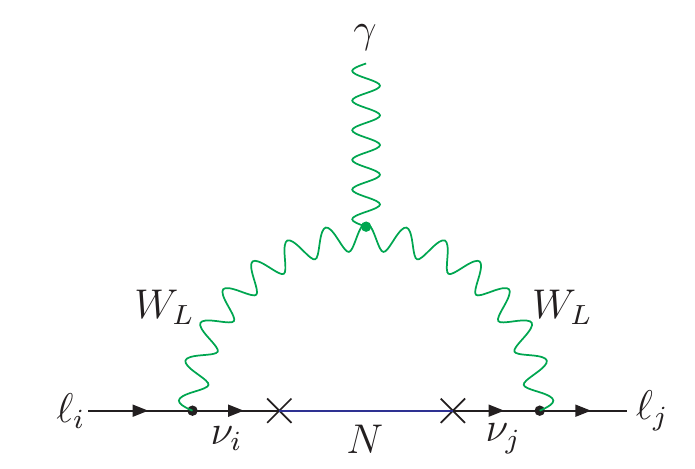} & 
\includegraphics[width=5cm]{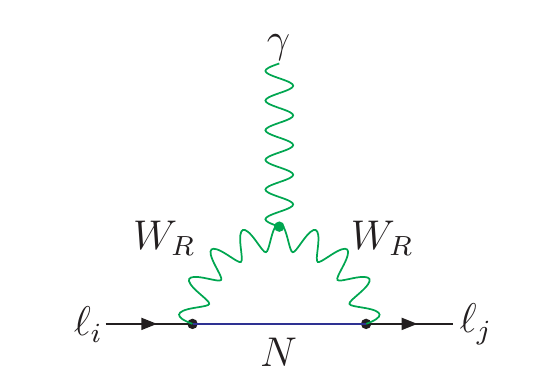} &
\includegraphics[width=5cm]{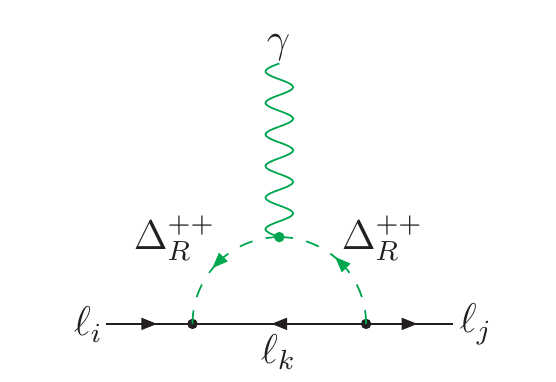} \\
(a) & (b) & (c) 
\end{tabular}
\end{center}
\caption{Various one-loop diagrams contributing to the LFV process $\ell_i\to \ell_j\gamma$ in the minimal L-R seesaw model.}\label{figmueg}
\end{figure}

In the minimal SM-seesaw, the only class of graphs that lead to enhanced LFV signal in $\mu\to e\gamma$ process arises from heavy-light mixing in second order and involve the $W_L$ exchange~\cite{mueg-L} (see Figure~\ref{figmueg}a). This contribution has been calculated most recently in Ref.~\cite{gavela} for cases where $M_D$ has special forms that lead to large heavy-light mixing $\xi$ and it has been noted that the branching ratio (BR) of $\mu\to e\gamma$ in this case can be as large as $10^{-13}$ for $M_N\leq 200$ GeV, whereas BR$(\tau\to\mu\gamma)$ can be as large as $10^{-9}$. All these are in the observable range of current and planned experiments~\cite{MEG, MEG2, prism}. 
However, as $M_N$ increases, these effects go down as its fourth power and become unobservable.

In the generic L-R model, the contributions to BR$(\mu\to e\gamma)$ and other charged LFV processes also receive contributions from new sources, e.g., $W_R$ exchange~\cite{riaz} and $\Delta^{++}_R$ exchange~\cite{rnm-92}, depending  on the details of the model. 
\subsubsection{$W_R$ contribution}
In the generic model, the $W_R-\nu_R$ virtual state gives a new contribution (see Figure~\ref{figmueg}b) as was noted in Ref.~\cite{riaz} which scales like $M^{-8}_{W_R}$ in the branching ratio:
\begin{eqnarray}
{\rm BR}({\mu\to e\gamma})_{W_R}\simeq \frac{3\alpha_W}{32\pi}\left(\frac{M_{W_L}}{M_{W_R}}\right)^8 \left(s_R c_R\frac{m^2_{N_2}-m^2_{N_1}}{M^2_{W_L}}\right)^2
\end{eqnarray}
where $\alpha_W=g^2/4\pi$ ($g$ being the weak coupling strength), and $s_R\equiv \sin\theta_R, c_R\equiv \cos\theta_R$, $\theta_R$ being the mixing angle in the RH charged current interaction in the lepton sector. The interesting aspect of this diagram is that it only depends on the mixing in the  RH charged current interaction with $W_R$  in a manner analogous to the well-known  GIM mechanism in the SM for the LH currents~\cite{gim}. In particular, it is independent of the observed neutrino mixing.  In our model due to the specific texture for the RH neutrinos given by Eq.~(\ref{eq:MN}), in the mass basis for charged leptons, we have an almost near degeneracy between the two lightest RH neutrinos that contribute to this process. As a result one finds that the $W_R$ contribution to BR$(\mu\to e\gamma)$
is many orders of magnitude smaller compared to the other contributions discussed below.
\subsubsection{Scalar contribution}
Another contribution to $\mu\to e\gamma$  comes from the scalar sector of the model involving $\Delta^{++}_R$ fields in the loop~\cite{rnm-92} (see Figure~\ref{figmueg}c):
\begin{eqnarray}
{\rm BR}({\mu\to e\gamma})_{\Delta_R^{++}}\simeq \frac{2\alpha_W M^4_{W_L}}{3\pi g^4}\left[\frac{(ff^\dagger)_{12}}{M^2_{\Delta_R^{++}}}\right]^2
\label{eq:brmuegd}
\end{eqnarray} 
This has been calculated for the `vanilla' L-R seesaw model under certain assumptions in Refs.~\cite{ciri}, and the current experimental limit on BR$(\mu\to e\gamma)$ implies a lower  bound on $M_{\Delta_R^{++}}\geq 1.7$ TeV for RH charged current mixing $\sim 0.01$. 
In our model, where all the RH neutrino masses and mixing angles are fixed by the neutrino mass fit discussed in Section~\ref{sec4}, we find that $ff^\dag$ is close to being diagonal, and hence, the contribution given by Eq.~(\ref{eq:brmuegd}) turns out to be very small. For the best-fit values given in Table~\ref{tab2}, we estimate this contribution to BR$({\mu\to e\gamma})$ to be $3.2\times 10^{-19}$ for $M_{\Delta_R^{++}} = 1.6$ TeV and $M_{W_R} = 3$ TeV.
\subsubsection{$W_L$ contribution}
\begin{figure}[t]
\centering
\includegraphics[width=8cm]{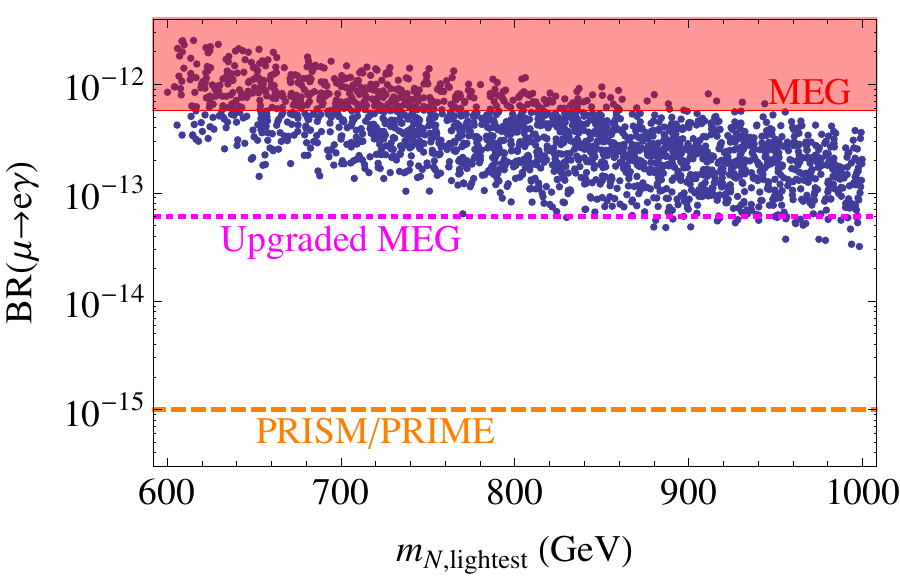}
\caption{Predictions for BR$(\mu\to e\gamma)$ in our TeV-scale L-R seesaw model for $M_{W_R}=3$ TeV. The (red) shaded region is excluded by the latest MEG results~\cite{MEG}. The (magenta) dotted line shows the sensitivity of the upgraded MEG experiment~\cite{MEG2}, and the (orange) dashed line shows the projected limit from the planned PRISM/PRIME experiment~\cite{prism}.}\label{scatmueg}
\end{figure}

This contribution, as evaluated in Ref.~\cite{mueg-L}, has the following form:
\begin{eqnarray}
{\rm BR}(\mu\to e\gamma)_{W_L} = \frac{\alpha_W^3s_W^2}{256\pi^2}\frac{m_\mu^4}{M_{W_L}^4}\frac{m_\mu}{\Gamma_\mu}|G_\gamma^{\mu e}|^2, \label{mueg}
\end{eqnarray}
where $s_W\equiv \sin\theta_W$ ($\theta_W$ being the weak mixing angle), $\Gamma_\mu=2.996\times 10^{-19}$ GeV~\cite{PDG} is the total decay width of the muon, and $G_\gamma^{\mu e}$ is the form factor given by 
\begin{eqnarray}
G_\gamma^{\mu e} &=& \sum_{i=1}^3 \xi_{ei}\xi^*_{\mu i} G_\gamma(x_{N_i})
\label{gmue}
\end{eqnarray}
where $x_{N_i}=m_{N_i}^2/M_{W_L}^2$, and 
\begin{eqnarray}
G_\gamma(x) &=& -\frac{x(2x^2+5x-1)}{4(1-x)^3}-\frac{3x^3}{2(1-x)^4}\ln x 
\end{eqnarray}
is the loop function which approximates to 1/2 for $x\gg 1$. 
It depends on the sum of the heavy-light mixing  parameters $\xi_{ei}\xi_{\mu i}$ which are predicted in our model. Due to the relative `largeness' of the mixing elements, this turns out to give the dominant contribution to BR$(\mu\to e\gamma)$ in our model, and restricts the allowed model parameter space significantly. As shown in Figure~\ref{scatmueg} and also in Table~\ref{tab5}, for the range of model parameter values given in Table~\ref{tab4}, the BR$(\mu\to e\gamma)$ predictions in our model are very close to the current MEG upper limit:  BR$(\mu\to e\gamma)\leq  5.7\times 10^{-13}$~\cite{MEG}. In fact, we find that the MEG limit has already ruled out part of the model parameter space which was otherwise consistent with all the leptonic masses and mixing. Most of the remaining parameter space shown here can therefore be tested once the BR$(\mu\to e\gamma)$ goes down by one order of magnitude in the upgraded MEG experiment~\cite{MEG2}. The full parameter space shown in Figure~\ref{scatmueg} is well within reach of the future high sensitivity searches such as PRISM/PRIME~\cite{prism}. The corresponding model predictions for charged LFV processes involving the third generation, namely, $\tau\to \mu \gamma$ and $\tau \to e\gamma$, are found to be much smaller than the current experimental sensitivity, as shown in Table~\ref{tab5}.  The above results are obtained for the choice of $M_{W_R}= 3 $ TeV (or $v_R\sim 5$ TeV), but can be easily scaled for higher values.
\begin{table*}[t]
\begin{center}
\begin{tabular}{c|c|c|c} \hline\hline
LFV Observable & Best-Fit Value & Range & Experimental Limit\\ \hline
BR($\mu \rightarrow e\gamma$) & $4.86\times 10^{-13}$ & ($3.2\times 10^{-14}$ - $2.6 \times \ 10^{-12}$) & $<5.7\times 10^{-13}$~\cite{MEG} \\
BR($\tau \rightarrow e\gamma$) & $1.08\times 10^{-12}$ & ($9.7\times 10^{-14}$ - $5.1 \times \ 10^{-12}$) & $<3.3\times 10^{-8}$~\cite{PDG} \\
BR($\tau \rightarrow \mu\gamma$) & $6.37\times 10^{-13}$ & ($3.7\times 10^{-14}$ - $3.7 \times \ 10^{-12}$) & $<4.4\times 10^{-8}$~\cite{PDG} \\ \hline
R$_{\mu \rightarrow e}^{\text{Ti}}$ & $4.26\times 10^{-13}$ & ($2.1\times 10^{-14}$ - $3.3\times 10^{-12}$) & $< 6.1 \times 10^{-13}$~\cite{mueTi}\\
R$_{\mu \rightarrow e}^{\text{Au}}$ & $3.80\times 10^{-13}$ & ($1.6\times 10^{-14}$ - $3.4\times 10^{-12} $) & $< 7.0 \times 10^{-13}$~\cite{mueAu} \\
R$_{\mu \rightarrow e}^{\text{Pb}}$ & $2.60\times 10^{-13}$ & ($1.1\times 10^{-14}$ - $2.4\times 10^{-12}$) &  $< 4.6 \times 10^{-11}$~\cite{muePb}\\
\hline\hline
\end{tabular}
\end{center}
\caption{The model predictions for various charged LFV processes $\ell_i\to \ell_j\gamma$ and $\mu\to e$ conversion in various nuclei. The best-fit and the range of values shown here correspond to those model parameter values shown in Tables~\ref{tab1} and \ref{tab4} respectively. The current experimental limits at 90\% CL are also shown for comparison.} \label{tab5}
\end{table*}
\subsection{$\mu \to e$ conversion in Nuclei} 
\begin{figure}[t]
\centering
\includegraphics[width = 8 cm]{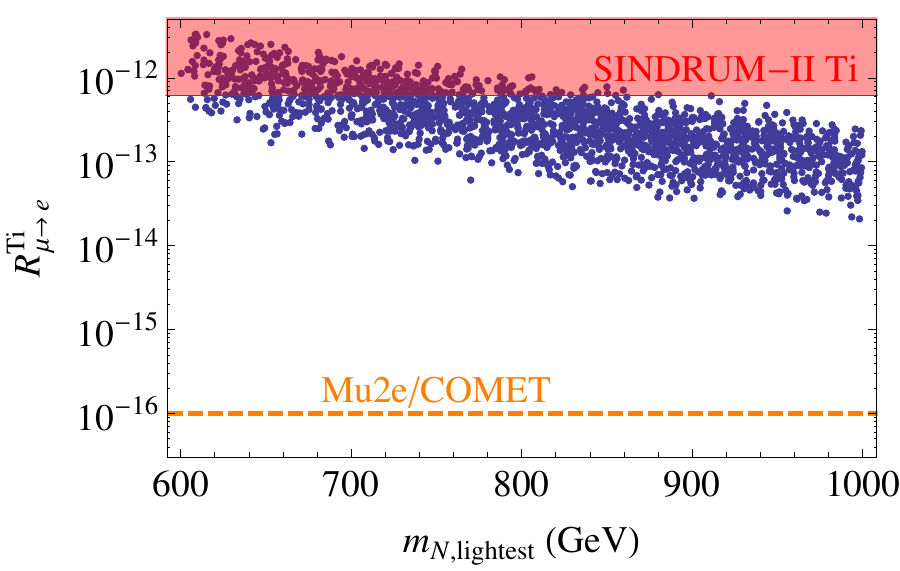} 
\hspace{0.1cm}
\includegraphics[width = 8 cm]{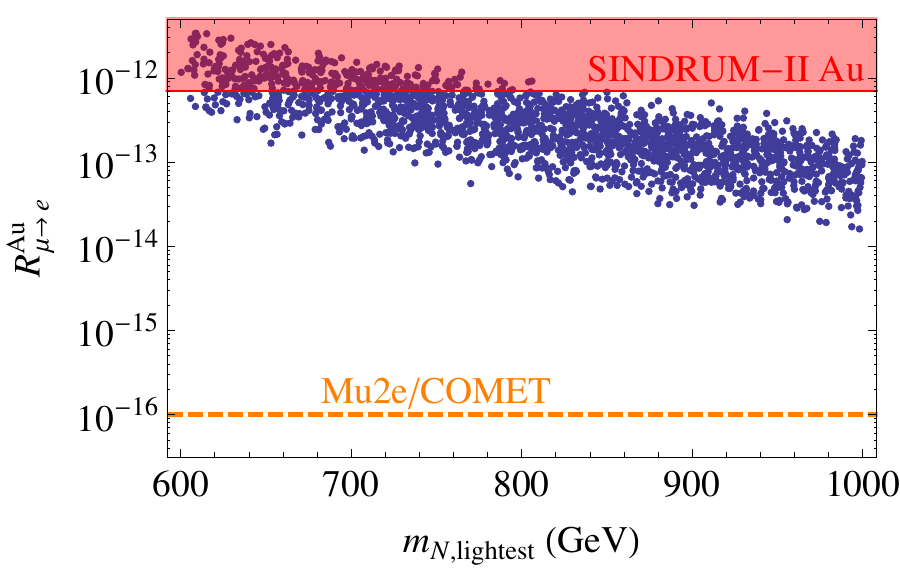}
\caption{Predictions for the $\mu\to e$ conversion rate on Titanium and Gold nuclei as a function of the mass of the lightest heavy neutrino in our TeV-scale L-R seesaw model. The (red) shaded region is excluded by SINDRUM-II results~\cite{mueTi, mueAu}, and the (orange) dashed line shows the future sensitivity of the proposed Mu2e/COMET experiment~\cite{mu2e}. }
\label{fig1}
\end{figure}

We also present the model predictions for $\mu \to e$ conversion in different nuclei. Again the $W_R$ contribution to this process is small compared to the $W_L$ contribution. We use the  expression for the ratio of $\mu\to e$ conversion rate over the capture rate from Ref.~\cite{gavela}:
\begin{eqnarray}
R_{\mu\to e} &=& \frac{2G_F^2\alpha_W^2 m_\mu^5}{16\pi^2 \Gamma_{\rm capt}}\left| 4V^{(p)}(2\tilde{F}_u^{\mu e}+\tilde{F}_d^{\mu e})
+4V^{(n)}(\tilde{F}_u^{\mu e}+2\tilde{F}_d^{\mu e})+s_W^2G_\gamma^{\mu e}\frac{D}{2e} \right|^2,
\label{mueconv}
\end{eqnarray}
where $G_F$ is the Fermi coupling constant, and $e$ is the magnitude of the electric charge. The dipole term $G_\gamma^{\mu e}$ is given in Eq.~(\ref{gmue}), and the form factors $\tilde{F}_{u,d}^{\mu e}$ are taken from Ref.~\cite{gavela} which in the limit $x_{N_i}\gg 1$ can also be written in the following compact form: 
\begin{eqnarray}
\tilde{F}_{u,d}^{\mu e} = \sum_{i=1}^3 \xi_{e i}\xi^*_{\mu i} \tilde{F}_{u,d}(x_{N_i})
\end{eqnarray}
with the loop functions
\begin{eqnarray}
\tilde{F}_u(x) &=& \frac{2}{3}s_W^2\frac{16\ln x - 37}{12}-\frac{3+3\ln x}{8},\\
\tilde{F}_d(x) &=& -\frac{1}{3}s_W^2\frac{16\ln x-37}{12}-\frac{3-3\ln x}{8}.
\end{eqnarray}
The nuclear form factors $D,V^{(p)}, V^{(n)}$, and the capture rate $\Gamma_{\rm capt}$ appearing in Eq.~(\ref{mueconv}) are summarized in Table~\ref{tabn} for various nuclei (the numbers were taken from Ref.~\cite{gavela}; see also Ref.~\cite{nucl}).  
\begin{table}[t]
\begin{center}
\begin{tabular}{c|c|c|c|c}\hline\hline
Nucleus ($_Z^A X$) & $V^{(p)}$ & $V^{(n)}$ & $D$ & $\Gamma_{\rm capt}~(10^6~{\rm sec}^{-1})$\\ \hline
$_{22}^{48}$Ti & 0.0396 & 0.0468 & 0.0864 & 2.59\\
$_{79}^{197}$Au & 0.0974 & 0.146 & 0.189 & 13.07\\
$_{82}^{208}$Pb & 0.0834 & 0.128 & 0.161 & 13.45\\
\hline\hline
\end{tabular}
\end{center}
\caption{The nuclear form factors and capture rates for various nuclei relevant for $\mu\to e $ conversion searches~\cite{gavela}. }\label{tabn}
\end{table}
The predicted ranges for our model parameter values given in Table~\ref{tab4} are depicted as a scatter 
plot in Figure~\ref{fig1} and also summarized in Table~\ref{tab5}. We find the conversion rates for Ti and Au nuclei to be only one/two orders of magnitude lower than the current upper 
limits~\cite{mueTi,mueAu}, and hence, can be probed in future high sensitivity searches~\cite{mu2e}. 

\begin{figure}[t]
\centering
\includegraphics[width=8cm]{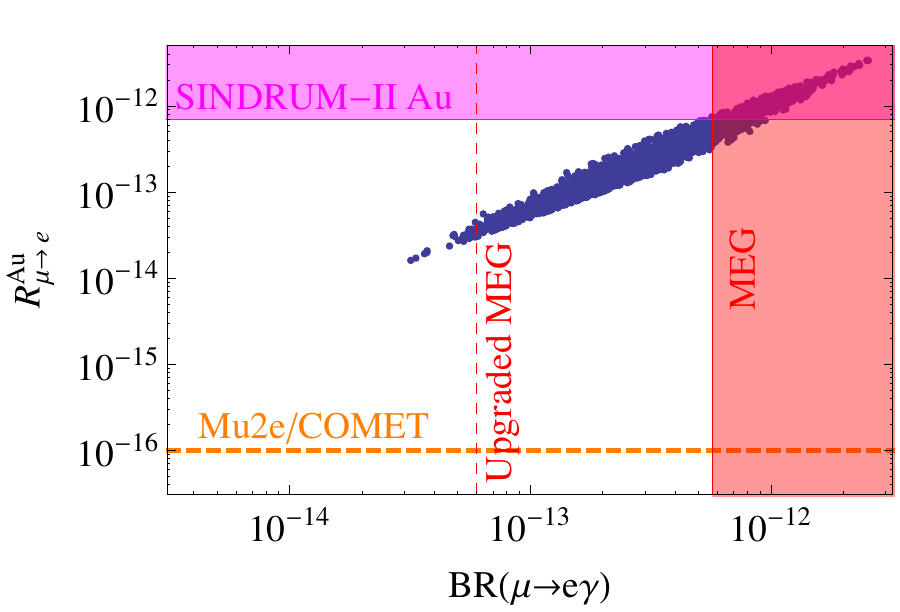}
\caption{The correlation between the $\mu\to e\gamma$ branching fraction and the $\mu\to e$ conversion rate in our model. The (red and magenta) shaded regions are excluded by MEG~\cite{MEG} and SINDRUM-II~\cite{mueAu} results respectively. The vertical and horizontal dashed lines show the projected limits on BR$(\mu\to e\gamma)$~\cite{MEG2} and $\mu\to e$ conversion rate~\cite{mu2e} respectively. }
\label{figcor}
\end{figure}

The correlation between the $\mu\to e\gamma$ branching fraction and the $\mu\to e$ conversion rate on muonic Gold, as predicted in our model, is shown  in Figure~\ref{figcor}. For other LFV branching ratios and conversion rates, similar correlations can be derived. The corresponding experimental limits and the future projections are also shown. Thus we find that the full model parameter space is accessible to the future high sensitivity searches for LFV. 
\subsection{$\mu\to 3 e$}
\begin{figure}[t]
\centering
\includegraphics[width=8cm]{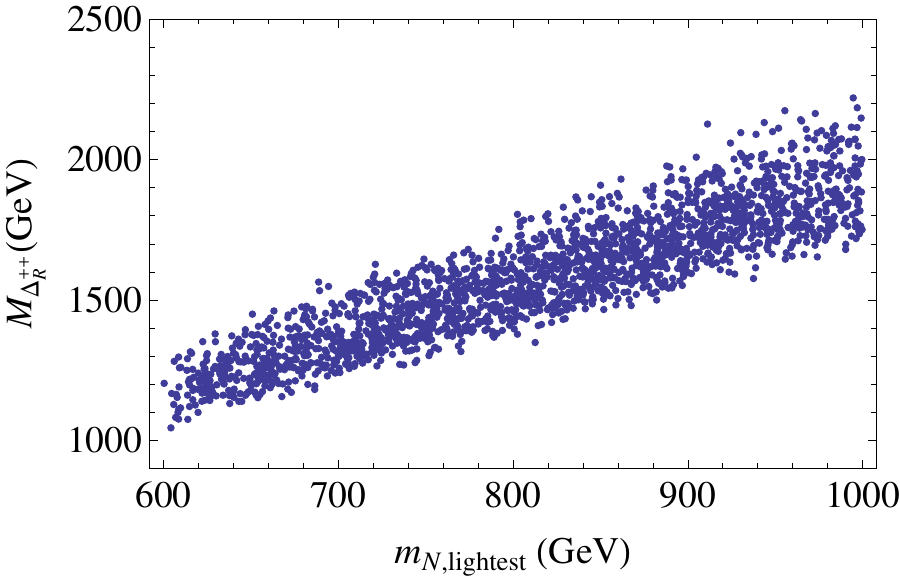}
\caption{The lower limit on the RH triplet mass as a function of the lightest heavy neutrino mass, as required to satisfy the BR$(\mu\to 3e)$ constraint.}\label{figmu3e}
\end{figure}

Turning now to another LFV process $\mu\to 3e$, again there are several contributions (see e.g., Ref.~\cite{gavela}): a photon-mediated one loop graph that contributes to $\mu\to e\gamma$ with a virtual $\gamma$-loop, box graphs with $W_R$ and $\nu_R$ virtual states plus a tree level graph involving the exchange of $\Delta^{++}_{R,L}$ states. The generic formula for the tree-level contribution is given by~\cite{pal}
\begin{eqnarray}
{\rm BR}({\mu\to 3e})\simeq \frac{1}{2}\left(\frac{M_{W_L}}{M_{W_R}}\right)^4\left(\frac{M'_{N,12}M'_{N,11}}{M^2_{\Delta^{++}_R}}\right)^2.
\end{eqnarray}
where $M'_{N,ij}$ are the elements of the RH neutrino mass matrix in the basis where the charged lepton mass matrix is diagonal (see Eq. (18)).
In our model, since the neutrino mass fit fixes all the parameters of the model except $M_{W_R}$ and $M_{\Delta_R^{++}}$, for a given $M_{W_R}$, we can find a lower limit on the RH triplet mass $M_{\Delta_R^{++}}$. For instance, for the best-fit parameter values given in Table~\ref{tab1} and for $M_{W_R}=3$ TeV, the current experimental upper limit on ${\rm BR}({\mu\to 3e})<1.0\times 10^{-12}$~\cite{PDG} implies a lower limit $M_{\Delta^{++}_R}>1.6$ TeV. For the range of model parameters given in Table~\ref{tab4}, the corresponding lower limits on $M_{\Delta_R^{++}}$ are shown in Figure~\ref{figmu3e}. 

\section{Leptonic Non-unitarity Effects}\label{sec6}
\begin{figure}[t]
\centering
\includegraphics[width=8cm]{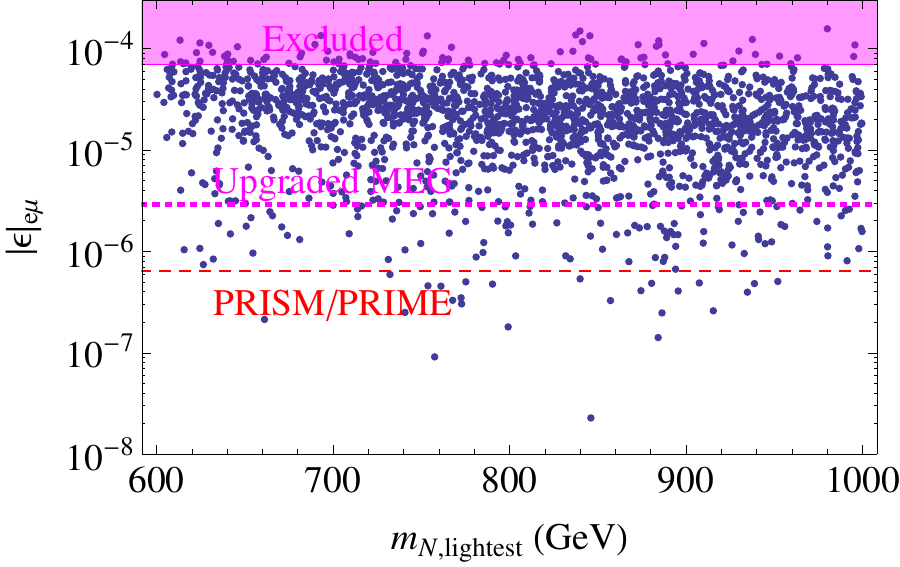}
\hspace{0.1cm}
\includegraphics[width=8cm]{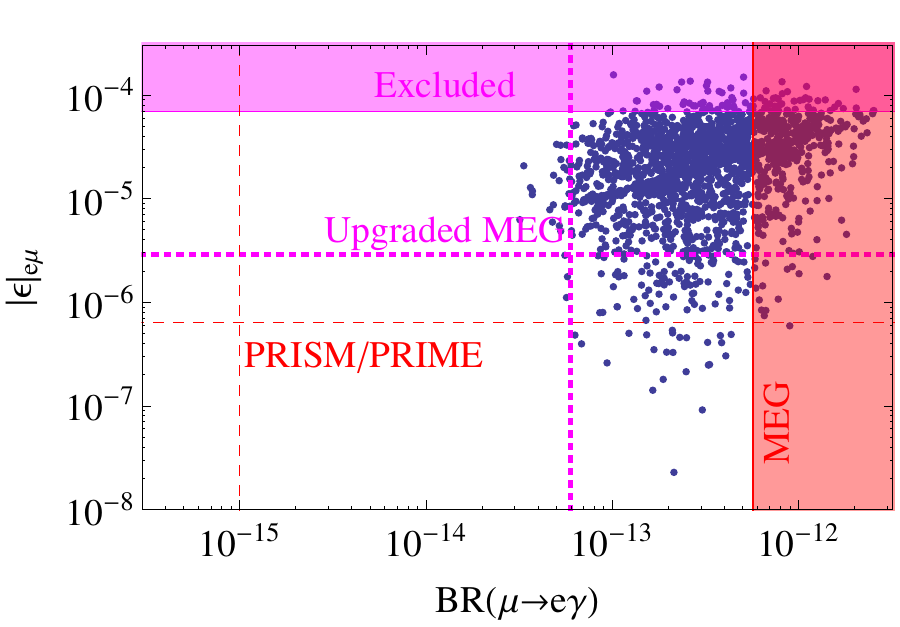}
\caption{Model predictions for the non-unitarity parameter $|\epsilon|_{e\mu}$ as a function of the lightest heavy neutrino mass (left panel) and its correlation with the BR$(\mu\to e\gamma)$ (right panel). The (magenta and red) shaded regions show the experimentally excluded regions, whereas the (magenta) dotted lines show the upgraded MEG sensitivity~\cite{MEG2}, and the (red) dashed lines show the projected sensitivity from PRISM/PRIME experiment~\cite{prism}.} \label{fig2}
 \end{figure}

In the usual 3-neutrino scenario, the $3\times 3$ light neutrino mass matrix is diagonalized by a unitary PMNS matrix. However, in the presence of heavy neutrinos, there are off-diagonal mixing between different light neutrinos due to their mixing with the heavy neutrinos which could be large in the special L-R models considered here. This leads to the non-unitarity of the light neutrino mixing matrix $U_L$, and can be parametrized by the deviation of $U_L^\dag U_L$ from the identity matrix. Using the best-fit values from Table~\ref{tab2}, we obtain
\begin{eqnarray}
\epsilon \equiv U_L^\dag U_L  = \left(\begin{array}{ccc}
0.99998 & 4.28\times 10^{-5} & -6.90\times 10^{-5}\\
4.28\times 10^{-5} & 0.99985 & 9.10\times 10^{-5}\\
-6.90\times 10^{-5} & 9.10\times 10^{-5} & 0.99987
\end{array}\right). 
\label{eq:nuty1}
\end{eqnarray}
For comparison, the 90\% CL limits on the non-unitarity parameters from a global fit of neutrino oscillation data, electroweak decays, universality tests, and rare charged lepton decays, are given by~\cite{antusch} (see also Ref.~\cite{Abada:2007ux})
\begin{eqnarray}
|\epsilon|_{\rm exp}  \approx
\left(\begin{array}{ccc}
0.994\pm 0.005 & <7.0\times 10^{-5} & <1.6\times 10^{-2}\\
<7.0\times 10^{-5} & 0.995\pm 0.005 & <1.0\times 10^{-2}\\
<1.6\times 10^{-2} & <1.0\times 10^{-2} & 0.995\pm 0.005
\end{array}\right),
\label{eq:nuty2}
 \end{eqnarray} 
all of which are respected by our best-fit values given in Eq.~(\ref{eq:nuty1}). For the full range of model  parameter values given in Table~\ref{tab4}, our predictions for various non-unitarity parameters are shown in Table~\ref{tab:nuty}. The predictions for the $|\epsilon|_{e\mu}$ element  and its correlation with the BR$(\mu\to e\gamma)$ are shown in Figure~\ref{fig2}, along with their current experimental limits as well as the projected limits. Most of the model parameter space is within reach of the projected limit: $|\epsilon|_{e\mu }<2.9\times 10^{-6}$ corresponding to the upgraded MEG limit on BR$(\mu\to e\gamma)<6\times 10^{-14}$~\cite{MEG2}.  For the PRISM/PRIME projected limit of BR$(\mu\to e\gamma)<10^{-15}$~\cite{prism}, the sensitivity for $|\epsilon|_{e\mu}$ can reach as low as $6.4\times 10^{-7}$. 
\begin{table}[t]
\begin{center}
\begin{tabular}{c|c|c|c} \hline\hline
Non-Unitarity Parameter & Best-Fit Value & Range & Experimental Limit~\cite{antusch}\\ \hline
$|\epsilon|_{e\mu}$ & $4.28\times 10^{-5}$ & $2.3\times 10^{-8}$ - $1.6\times 10^{-4}$ & $<7.0\times 10^{-5}$ \\
$|\epsilon|_{e\tau}$ & $6.90\times 10^{-5}$ & $1.6\times 10^{-7}$ - $2.2\times 10^{-4}$ & $<1.6\times 10^{-2}$ \\
$|\epsilon|_{\mu\tau}$ & $9.10\times 10^{-5}$ & $2.2\times 10^{-8}$ - $4.1\times 10^{-4}$& $<1.0\times 10^{-2}$ \\
\hline\hline
\end{tabular}
\end{center}
\caption{The model predictions for the non-unitarity parameters. The best-fit and the range of values shown here correspond to those model parameter values shown in Tables~\ref{tab1} and \ref{tab4} respectively. The current experimental limits at 90\% CL are also shown for comparison.}
\label{tab:nuty}
\end{table}
\section{Implications for neutrinoless double beta decay}\label{sec7}
\begin{figure}[t]
\begin{center}
\begin{tabular}{ccc}
\includegraphics[width=5cm]{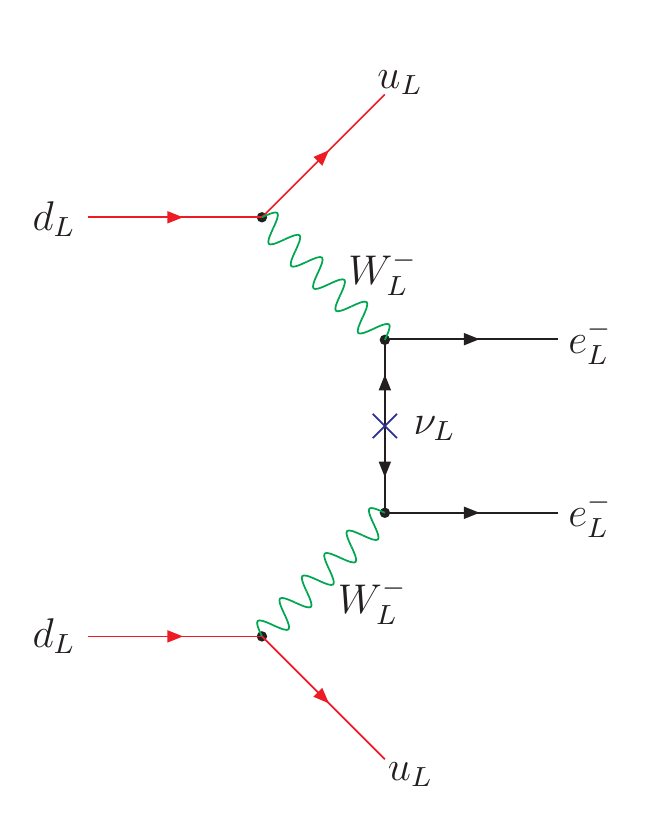} & 
\includegraphics[width=5cm]{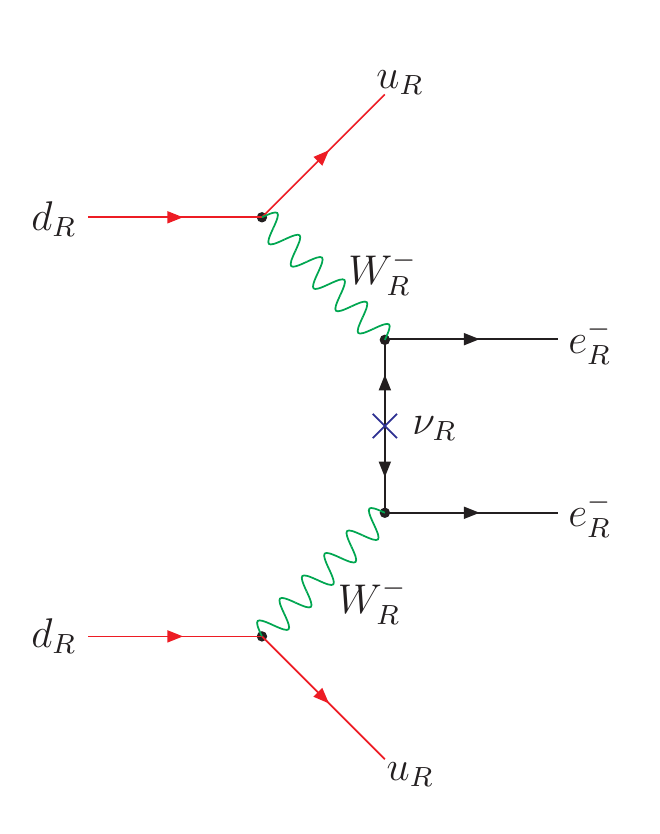} &
\includegraphics[width=5cm]{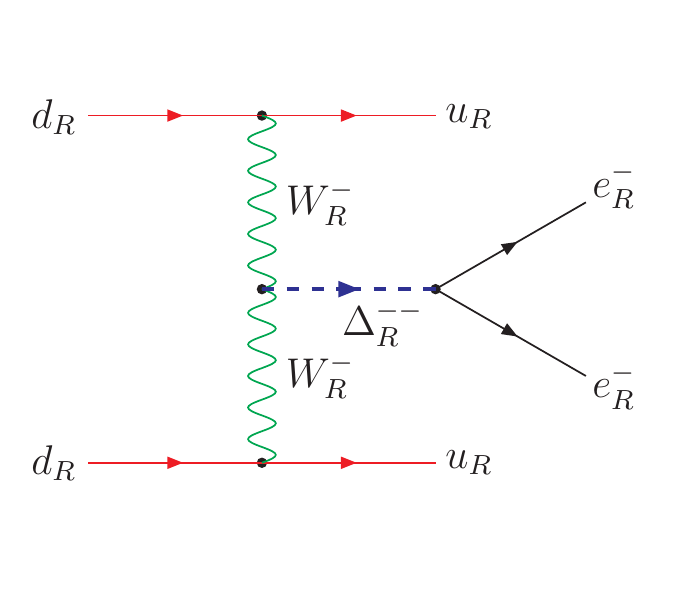} \\
(a) & (b) & (c) \\ 
\includegraphics[width=5cm]{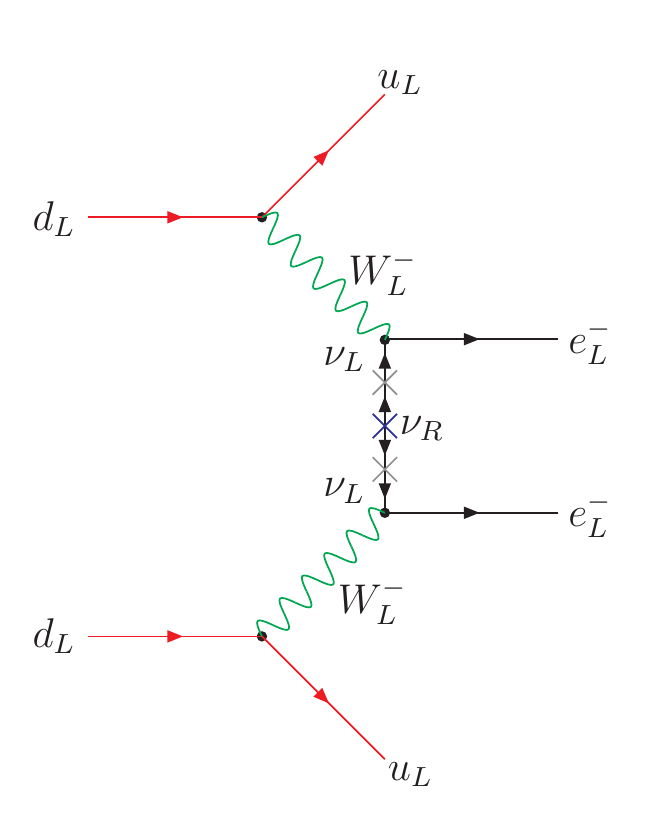} & 
\includegraphics[width=5cm]{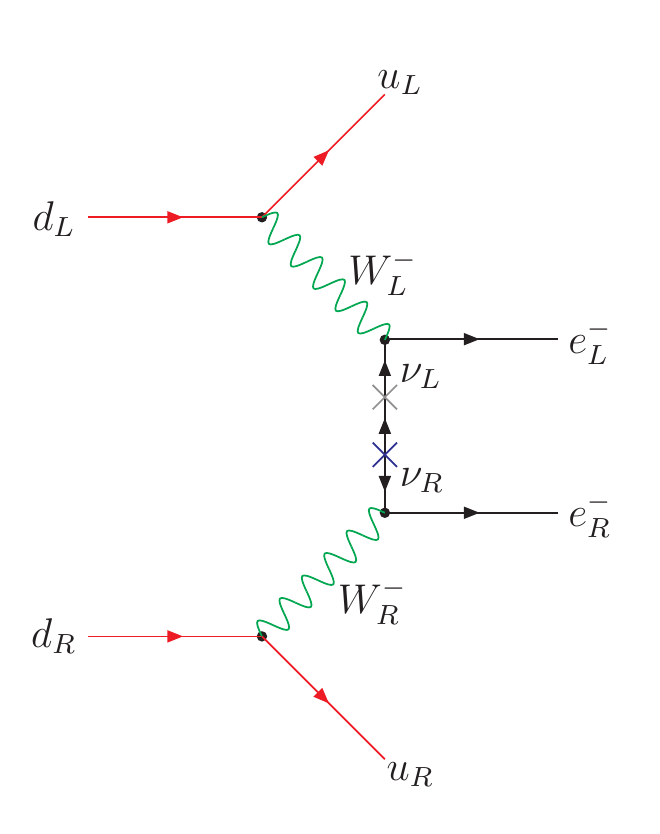} &
\includegraphics[width=5cm]{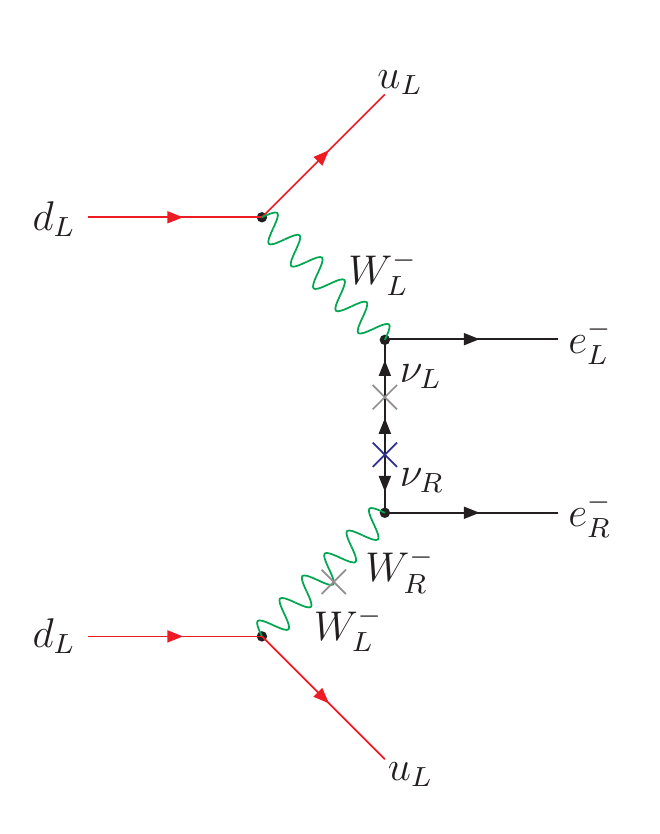} \\
(d) & (e) & (f) 
\end{tabular}
\end{center}
\caption{The dominant Feynman diagrams contributing to the $0\nu\beta\beta$ process in our L-R seesaw 
model, due to (a) light neutrino exchange with purely LH current, (b) heavy neutrino exchange with purely RH current, (c) scalar triplets with RH current, (d) heavy neutrino exchange with LH current, (e) heavy-light neutrino mixing ($\lambda$-diagram), (f) heavy-light neutrino and $W_L-W_R$ mixing ($\eta$-diagram).}\label{figdb}
\end{figure}
In this section, we discuss tests of the TeV L-R seesaw model in $0\nu\beta\beta$ process. Since in L-R seesaw models both the light ($\nu_{L,e}$) and heavy ($\nu_{R,e}$) neutrinos  are Majorana fermions, they break lepton number by two units and lead to the $0\nu\beta\beta$ process $^A_ZX\to ^A_{Z+2}Y+e^-+e^-$~\cite{rode}. The first contribution to this process comes from the well-known light neutrino exchange via LH current (see Figure~\ref{figdb}a) whose amplitude is given by $A_{\nu} \simeq G_F^2m_{ee}/q^2$, 
where $|q^2|\sim (100~{\rm MeV})^2$ is the typical momentum exchange scale, and $m_{ee}\equiv |\sum_i U_{L,ei}^2 m_{\nu,i}|$ is the so-called effective mass. This is usually expressed in terms of the dimensionless parameter 
\begin{eqnarray}
|\eta_\nu^L| \equiv \left|\frac{m_{ee}}{m_e}\right| = \left|\frac{\sum_i U_{L,ei}^2 m_{\nu,i}}{m_e}\right|,
\label{eta1}
\end{eqnarray}
where $m_e$ is the electron mass. 
In the L-R model, there is an analogous contribution from the RH sector involving heavy neutrinos and $W_R$~\cite{MS} (see Figure~\ref{figdb}b), which can be parametrized by 
\begin{eqnarray}
|\eta_{\nu_R}^R| = m_p\left(\frac{M_{W_L}}{M_{W_R}}\right)^4\left|\sum_i \frac{V_{R, ei}^2}{m_{N_i}}\right|,
\label{eta2}
\end{eqnarray}
where $m_p$ is the mass of the proton. 
There are also other contributions of similar order of magnitude coming from $\Delta^{--}_R$ exchange~\cite{MV} (see Figure~\ref{figdb}c), which is parametrized by 
\begin{eqnarray}
|\eta_{\Delta_R}| = \frac{m_p}{G_F^2}\left|\frac{\sum_i V_{R, ei}^2 m_{N_i}}{M_{\Delta_R^{--}}^2M_{W_R}^4}\right|.
\label{eta3}
\end{eqnarray}
For large heavy-light neutrino mixing and/or $W_L-W_R$ mixing, there are additional contributions due to mixed LH-RH currents~\cite{hirsch}, and must also be taken into account. The dominant mixed contributions come from the so-called $\lambda$ and $\eta$ diagrams (Figures~\ref{figdb}e, \ref{figdb}f):
\begin{eqnarray}
|\eta_\lambda| &=& \left(\frac{M_{W_L}}{M_{W_R}}\right)^2\left|\sum_i U_{L,ei}{\xi'}^*_{ei}\right|,\label{eta4}\\
|\eta_\eta| &=& \tan\zeta \left|\sum_i U_{L,ei}{\xi'}^*_{ei}\right|,
\label{eta5}
\end{eqnarray}
where $\tan2\zeta\lsim (M_{W_L}/M_{W_R})^2$ represents the mixing between the LH and RH gauge bosons, and $\xi'$ is defined by Eq.~(\ref{eq:big}). There is also a diagram in which the heavy neutrinos are exchanged with purely LH currents (Figure~\ref{figdb}d), and this is parametrized by
\begin{eqnarray}
|\eta_{\nu_R}^L| = m_p\left|\sum_i \frac{\xi_{ei}^2}{m_{N_i}}\right|.
\label{eta6}
\end{eqnarray}
There is an analogous diagram in which the light neutrinos are exchanged with RH currents; however this is highly suppressed compared to the canonical one with LH currents. 

\begin{table}[t]
\begin{center}
\begin{tabular}{c|c|c}\hline\hline
Parameter & Value & Current Limit~\cite{barry} \\ \hline
$|\eta_\nu^L|$ & $8.1\times 10^{-11}$ & $\lsim 7.1\times 10^{-7}$\\
$|\eta_{\nu_R}^R|$ & $4.4\times 10^{-12}$ & $\lsim 7.0\times 10^{-9}$\\
$|\eta_{\nu_R}^L|$ & $1.2\times 10^{-19}$ & $\lsim 7.0\times 10^{-9}$\\
$|\eta_{\Delta_R}|$ & $2.1\times 10^{-10}$ & $\lsim 7.0\times 10^{-9}$\\
$|\eta_\lambda|$ & $1.5\times 10^{-8}$ & $\lsim 5.7\times 10^{-7}$\\
$|\eta_\eta|$ & $1.5\times 10^{-9}$ & $\lsim 3.0\times 10^{-9}$\\ 
\hline \hline
\end{tabular}
\end{center}
\caption{Model predictions for various contributions to the $0\nu\beta\beta$ process in our L-R model, corresponding to the best-fit values in Table~\ref{tab2}. Here we have taken $M_{W_R}=3$ TeV, $M_{\Delta_R^{--}}=1.6$ TeV and $\tan\zeta=0.1(M_{W_L}/M_{W_R})^2$. }\label{tab6}
\end{table}
The predictions for various contributions to the $0\nu\beta\beta$ amplitude discussed above for  our best-fit model parameters (as shown in Table~\ref{tab2}) are given in Table~\ref{tab6}. We have also shown the current limits on these parameters~\cite{barry} derived using the recent KamLAND-Zen results~\cite{kamland}. Here we have chosen $M_{W_R}=3$ TeV, $M_{\Delta_R^{++}}=1.6$ TeV (to be consistent with the $\mu\to 3e$ limit), and $\tan\zeta=0.1 (M_{W_L}/M_{W_R})^2$. Note that in our model with non-negligible LH-RH neutrino mixing, the $\eta$-diagram gives the dominant contribution to the $0\nu\beta\beta$ process for the above choice of parameters. In particular, for a given $W_R$ mass, we can derive an upper limit on the ratio of the vevs of bi-doublet $\phi_0$ to be $\delta\kappa/\kappa'\lsim 0.1$ which determines the $W_L-W_R$ mixing, and hence, the size of the $\eta$-contribution in Eq.~(\ref{eta5}). 
\begin{table}[t]
\begin{center}
\begin{tabular}{c|c|c|c|c|c}\hline\hline
Nucleus & $G_{01}^{0\nu}~({\rm yr}^{-1})$ & ${\cal M}_{\nu}^{0\nu}$ &  ${\cal M}_{\nu_R}^{0\nu}$ & ${\cal M}_{\lambda}^{0\nu}$ & ${\cal M}_{\eta}^{0\nu}$ \\ \hline
$^{76}$Ge & $5.77\times 10^{-15}$ & 2.81 - 5.82& 232 - 411.5 & 1.75 - 3.76 & 235 - 637 \\
$^{136}$Xe & $3.56\times 10^{-14}$  & 1.89 - 4.20 & 159.7 - 172.1 & 1.96 - 2.49 & 370 - 419
\\ \hline\hline
\end{tabular}
\end{center}
\caption{The $0\nu\beta\beta$ phase space factor and the nuclear matrix elements for two different nuclei. The phase space factors are taken from Ref.~\cite{kotilla} for the axial-vector coupling constant $g_A=1.25$. The range of NMEs for light neutrino exchange (${\cal M}_{\nu}^{0\nu}$) was taken from the compilation in Ref.~\cite{DGMR}, whereas for heavy neutrino exchange (${\cal M}_{\nu_R}^{0\nu}$), the range was taken from Ref.~\cite{nme1}. The $\lambda$ and $\eta$ NMEs were taken from the old calculations in Ref.~\cite{nme2}.} \label{tab:nme}
\end{table}

In order to compare our $0\nu\beta\beta$ predictions directly with experimental values, we translate the $\eta$'s defined above to the half-life: 
\begin{eqnarray}
\frac{1}{T_{1/2}^{0\nu}} &=& G_{01}^{0\nu}\left[|{\cal M}_\nu^{0\nu}|^2|\eta_{\nu}^L|^2
+|{\cal M}_{\nu_R}^{0\nu}|^2(|\eta_{\nu_R}^L|^2+|\eta_{\nu_R}^R+\eta_{\Delta_R}|^2)
+|{\cal M}_{\lambda}^{0\nu}|^2|\eta_\lambda|^2
+|{\cal M}_{\eta}^{0\nu}|^2|\eta_\eta|^2\right.\nonumber\\
&&\left.~~~~~+~{\rm interference~terms}\right],
\label{eq:half}
\end{eqnarray}
where $G_{01}^{0\nu}$ is the phase space factor, and ${\cal M}^{0\nu}$'s are the nuclear matrix elements (NMEs). 
The values of the phase space factor and the NMEs used in Eq.~(\ref{eq:half}) are given in Table~\ref{tab:nme} for two different nuclei $^{76}$Ge and $^{136}$Xe. The model predictions for the corresponding half-lives of $0\nu\beta\beta$ process are shown in Table~\ref{tab7}. The predicted ranges are due to the model parameter ranges in Table~\ref{tab4} and also due to the NME uncertainties in Table~\ref{tab:nme}. We also show the current experimental limits for $^{136}$Xe from KamLAND-Zen (and the combined limit from KamLAND-Zen+EXO-200)~\cite{kamland}  as well as for $^{76}$Ge from GERDA phase-I (and the combined limit from GERDA+Heidelberg-Moscow+IGEX)~\cite{gerda}. The future limit on $T_{0\nu}^{1/2}(^{76}{\rm Ge})$~\cite{gerda3, majorana} and $T_{0\nu}^{1/2}(^{136}{\rm Xe})$~\cite{exo1000} can probe the full model parameter space shown here. Note that the model predictions given here are for the conservative values $M_{W_R}=3$ TeV, $M_{\Delta_R^{++}}=1.6$ TeV and $\tan\zeta=0.1(M_{W_L}/M_{W_R})^2$, and hence, can only serve as the lowest representative values. 
\begin{table}[t]
\begin{center}
\begin{tabular}{c|c|c|c}\hline\hline
Nucleus & Model Prediction for $T_{1/2}^{0\nu}$ (yr) & Current Limit (yr) & Future Limit (yr)\\ \hline
$^{76}$Ge & $6.2\times 10^{25}$ - $6.2\times 10^{27}$ & $>2.1~ (3.0) \times 10^{25}$ ~\cite{gerda} & $6\times 10^{27}$~\cite{gerda3, majorana}\\
$^{136}$Xe & $2.3\times 10^{25}$ - $4.3\times 10^{26}$ & $>1.9~ (3.1) \times 10^{25}$~\cite{kamland} & $8\times 10^{26}$~\cite{exo1000}\\
\hline\hline
\end{tabular}
\end{center}
\caption{The model predictions for the $0\nu\beta\beta$ half-lives corresponding to the model parameters given in Table~\ref{tab4}, and for $M_{W_R}=3$ TeV, $M_{\Delta_R^{--}}=1.6$ TeV and $\tan\zeta=0.1(M_{W_L}/M_{W_R})^2$. The shown range also includes the NME uncertainties given in Table~\ref{tab:nme}. Also shown are the current experimental limits from GERDA~\cite{gerda} and KamLAND-Zen~\cite{kamland}. The numbers in parenthesis show the combined limits from GERDA+Heidelberg-Moscow+IGEX~\cite{gerda} and KamLAND-Zen+EXO-200~\cite{kamland}. The 
future projected limits for both isotopes~\cite{gerda3, majorana, exo1000} are also shown for comparison. }\label{tab7}
\end{table} 
\section{Collider Signature}\label{sec8}
\begin{figure}[t]
\begin{center}
\begin{tabular}{ccc}
\includegraphics[width=5cm]{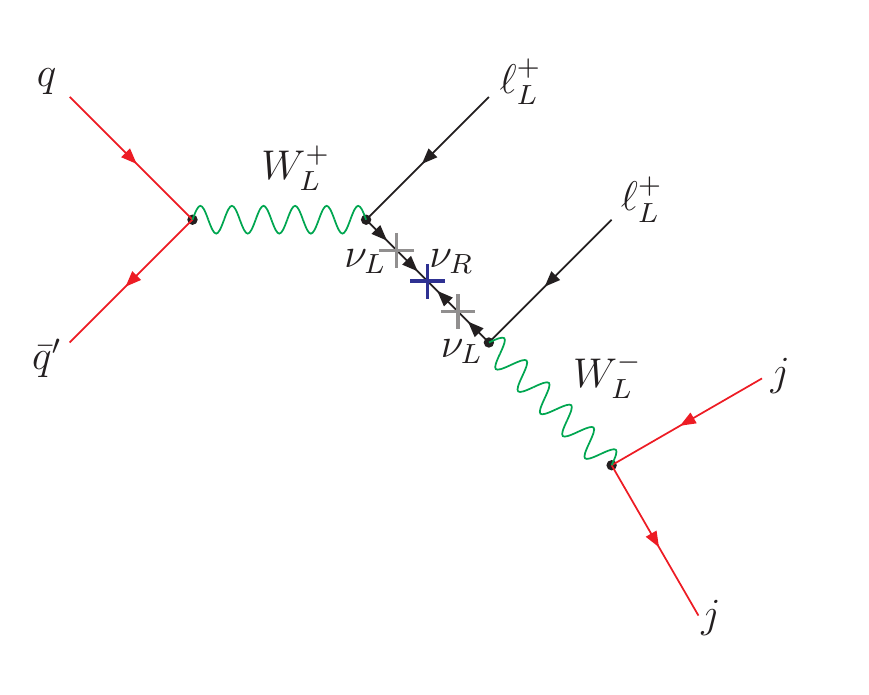} & 
\includegraphics[width=5cm]{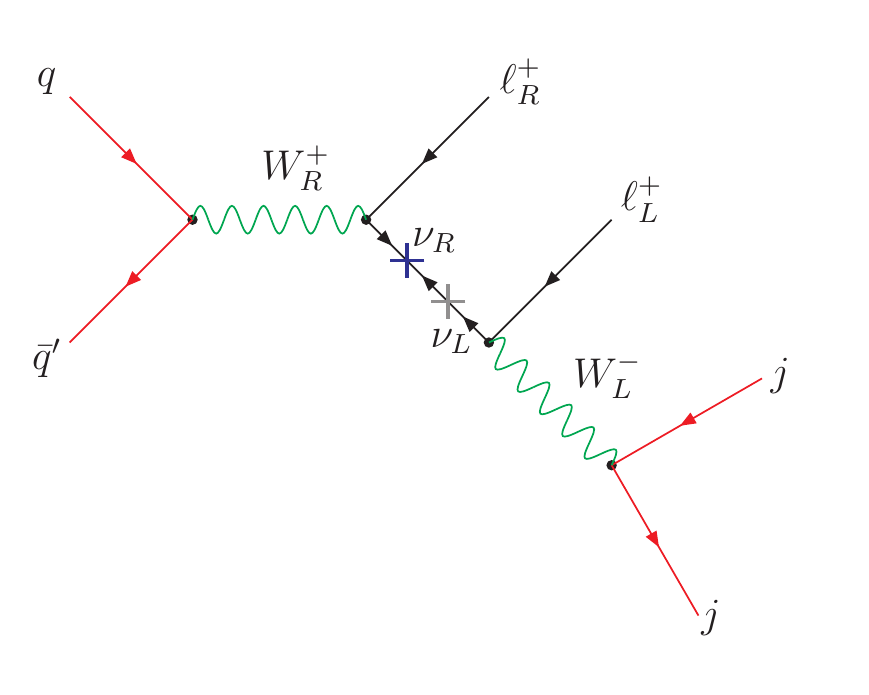} &
\includegraphics[width=5cm]{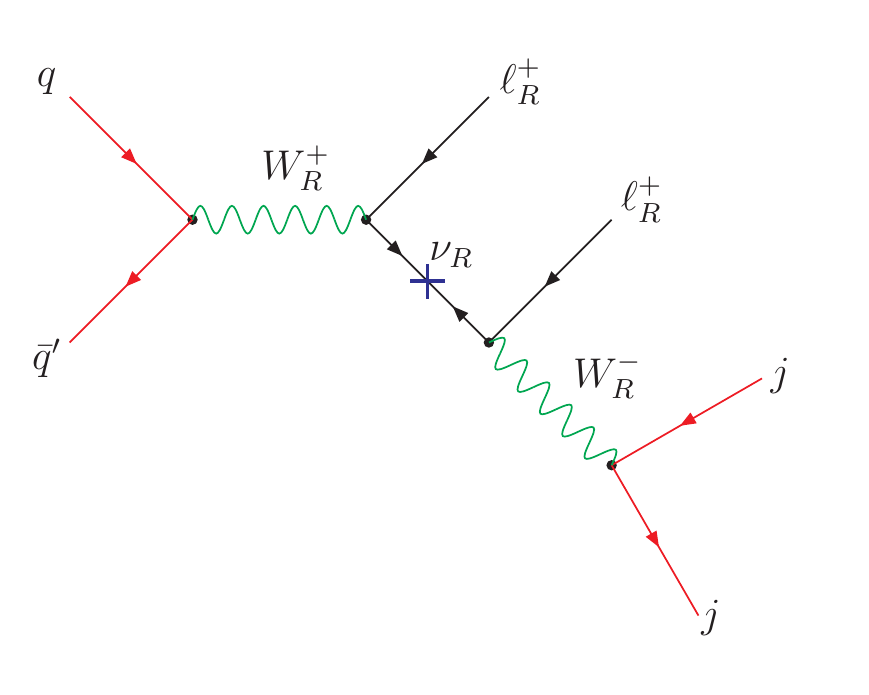} \\
(a) & (b) & (c) 
\end{tabular}
\end{center}
\caption{The dominant Feynman diagrams contributing to the smoking gun collider signal $\ell^\pm\ell^\pm jj$ of a heavy Majorana neutrino in our L-R seesaw model: (a) $LL$ channel with purely LH current, (b) $RL$ channel with mixed RH-LH current, (c) $RR$ channel with purely RH current.}\label{figcol}
\end{figure}

Both the Majorana nature and heavy-light mixing could manifest simultaneously via their distinct collider signals, thus giving complementary information to what is obtained from the low energy searches. There exist limits on the seesaw parameters $M_N$ and $\xi_{\ell i}$ for $M_N<M_Z$ from LEP electroweak precision 
data (for a review, see Ref.~\cite{atre}). With the recent LHC Higgs data, these limits have been extended to $M_N\leq 200$ GeV~\cite{dfm}. 
As far as direct collider tests of generic type-I seesaw is concerned, the smoking gun signal is the same-sign dileptons plus two jets without missing energy ($\ell^\pm\ell^\pm jj$)~\cite{KS}. In the absence of L-R symmetry, we need suitable combinations of sub-TeV Majorana mass $M_N$ as well as enhanced mixing $\xi_{\ell i}$ to have an observable signal at the LHC~\cite{theory-LL}. An important point is that it crucially depends on the ``largeness" of the heavy-light neutrino mixing, as the amplitude for this process (Figure~\ref{figcol}a) depends on the square of the mixing parameter $\xi$. The current LHC limits on $|\xi_{\ell i}|^2$ (for $\ell=e,\mu$) is between 0.01 - 0.1 for $M_N$=100 - 300 GeV~\cite{CMS-LL}. Including infrared enhancement effects~\cite{DPY}, these limits can be improved by at least a factor of 5 at $\sqrt s=14$ TeV. 

In the L-R symmetric embedding of TeV scale seesaw, the presence of RH gauge interactions could lead to significant enhancement for $\ell^\pm\ell^\pm jj$ signal, from $W_R$-mediated production and decay of $\nu_R$ (Figure~\ref{figcol}c) as was first pointed out in Ref.~\cite{KS} (for some recent LHC studies of this channel, see Refs.~\cite{theory-RR}). Two significant differences emerge in our model when compared to the signature in Ref.~\cite{KS}:

 (i) As recently noted there will be a new contribution (see Figure~\ref{figcol}b) to the like-sign dilepton signal from the fact that the heavy-light neutrino mixing  $\xi_{\ell i}$ in our model can be large \cite{chendev}. This is in addition to the usual $W_R$-mediated (Figure~\ref{figcol}c) and $W_L$-mediated  (Figure~\ref{figcol}a) diagrams. Note that the amplitude for the $RR$ diagram in Figure~\ref{figcol}c is independent of the heavy-light mixing, and hence, does not probe the full seesaw matrix. On the other hand, as shown in Ref.~\cite{chendev}, the $RL$ diagram is sensitive to the heavy-light mixing, and in fact, the dominant channel over a fairly large range of model parameter space. Following the  theoretical arguments based on vacuum stability which suggest that the heavy neutrinos in the minimal L-R seesaw models are lighter than the RH gauge bosons~\cite{rnm}, it was shown~\cite{chendev} that for RH gauge boson masses below 5 - 6 TeV, when it can be produced at the $\sqrt s=14$ TeV LHC with an observable cross section~\cite{ferrari}, its direct decay to on-shell RH neutrinos, which subsequently decay to the SM $W_L$-boson and charged leptons (Figure~\ref{figcol}b), will allow a probe of the heavy-light neutrino mixing for a wider mass range of up to a few TeVs from a study 
of the $\ell^\pm\ell^\pm jj$ signal. This new channel could also be useful in distinguishing the RH gauge boson contributions to the collider signatures from the LH ones~\cite{chendev, Chen:2011hc}, which provides complementary information to that obtained from low-energy experiments in the exploration of the low-scale L-R symmetry~\cite{snowmass}.  

(ii) Secondly, due to the specific texture of the RH neutrino mass matrix in our particular model, both the $W_R$ and $\xi_{\ell i}$ contributions will lead  predominantly to 
$\mu^\pm e^\pm jj$ final states which probe both lepton number violation and lepton flavor violation at the LHC. In fact, it is interesting to note that the same RH neutrino texture embedded into the SM seesaw will not lead to any like-sign dilepton signal.{\footnote{For quasi-degenerate Majorana neutrinos (as usually the case with specific textures), the lepton number violating collider signal in the SM seesaw is negligible except when the mass splitting between them is comparable to their width~\cite{Bray:2007ru}. For the textures in Eq.~(\ref{eq:texture}), the mass splitting turns out to be much smaller than the width, thus suppressing the $LL$ signal. But the new $RL$ signal is still large in our case. }
\section{Conclusion}\label{sec9}
We have presented a new TeV-scale seesaw model based on the left-right symmetric gauge group but without parity symmetry where a particular texture for the Dirac and Majorana masses guarantees that neutrino masses are naturally small while keeping the heavy-light neutrino mixing in the LHC-observable range. A discrete 
flavor symmetry is shown to guarantee the stability of this texture, while being consistent with the observed lepton masses and mixing. We then explore its tests in the domain of charged lepton flavor violating processes such as $\mu\to e\gamma$ and $\mu\to e$ conversion. We find that if the $SU(2)_R$-breaking scale is of order 5 TeV, it can be tested in near future when the experimental sensitivity for the branching ratios of $\mu\to e\gamma$ as well as $\mu\to e$ conversion are improved by an order of magnitude below the current limits. We also present predictions for leptonic non-unitarity and its correlations with the BR$(\mu\to e\gamma)$. In addition, we also study the neutrinoless double beta decay for the model parameters in the text. It is interesting that parts of the predicted range are also accessible to current and future 
double beta decay searches. Finally, we briefly discuss various contributions to the smoking gun collider signature in this model. As noted in the text, our results for both lepton flavor violation and lepton number violation can easily be extended for higher values of the $SU(2)_R$-scale without affecting the neutrino fits presented here. The model also predicts sub-TeV SM-like Higgs bosons which are leptophilic with rich 
collider phenomenology, and this is currently under investigation.

\section*{Acknowledgments}
We would like to acknowledge useful discussions with K. S. Babu, F. del Aguila, T. Hambye, A. Pilaftsis, G. Senjanovi\'c and V. Tello. PSBD acknowledges the local support and hospitality provided by the Maryland Center for Fundamental Physics during the initial stages of this work. The work of PSBD is supported by the 
Lancaster-Manchester-Sheffield Consortium for Fundamental Physics under STFC grant ST/J000418/1. CHL and RNM are supported by the National Science Foundation grant No. PHY-0968854.

\end{document}